\documentclass[english,aps,pra,preprint,groupedaddress,nofootinbib]{revtex4}
\usepackage[T1]{fontenc}
\usepackage[latin9]{inputenc}
\usepackage{amstext}
\usepackage{graphicx}
\usepackage{esint}

\makeatletter

\providecommand*{\perispomeni}{\char126}
\AtBeginDocument{\DeclareRobustCommand{\greektext}{%
  \fontencoding{LGR}\selectfont\def\encodingdefault{LGR}%
  \renewcommand{\~}{\perispomeni}%
}}
\DeclareRobustCommand{\textgreek}[1]{\leavevmode{\greektext #1}}
\DeclareFontEncoding{LGR}{}{}

\@ifundefined{textcolor}{}
{%
 \definecolor{BLACK}{gray}{0}
 \definecolor{WHITE}{gray}{1}
 \definecolor{RED}{rgb}{1,0,0}
 \definecolor{GREEN}{rgb}{0,1,0}
 \definecolor{BLUE}{rgb}{0,0,1}
 \definecolor{CYAN}{cmyk}{1,0,0,0}
 \definecolor{MAGENTA}{cmyk}{0,1,0,0}
 \definecolor{YELLOW}{cmyk}{0,0,1,0}
 }

\makeatother

\begin{document}

\title{Classical phases and quantum angles in the description of interfering
Bose-Einstein condensates}

\author{W. J. Mullin$^{a}$ and F. Laloë$^{b}$}

\affiliation{$^{a}$Department of Physics, University of Massachusetts, Amherst,
Massachusetts 01003 USA\\
 $^{b}$Laboratoire Kastler Brossel, ENS, UPMC, CNRS; 24 rue Lhomond,
75005 Paris, France}

\email{mullin@physics.umass.edu;laloe@lkb.ens.fr}
\begin{abstract}
The interference of two Bose-Einstein condensates, initially in Fock
states, can be described in terms of their relative phase, treated as
a random unknown variable.  This phase can be understood, either as
emerging from the measurements, or preexisting to them; in the latter
case, the originating states could be phase states with unknown
phases, so that an average over all their possible values is taken.
Both points of view lead to a description of probabilities of results
of experiments in terms of a phase angle, which plays the role of a
classical variable.  Nevertheless, in some situations, this
description is not sufficient: another variable, which we call the
{}``quantum angle'', emerges from the theory.  This article studies
various manifestations of the quantum angle.  We first introduce the
quantum angle by expressing two Fock states crossing a beam splitter
in terms of phase states, and relate the quantum angle to off-diagonal
matrix elements in the phase representation.  Then we consider an
experiment with two beam splitters, where two experimenters make
dichotomic measurements with two interferometers and detectors that
are far apart; the results lead to violations of the
Bell-Clauser-Horne-Shimony-Holt inequality (valid for local-realistic
theories, including classical descriptions of the phase).  Finally, we
discuss an experiment where particles from each of two sources are
either deviated via a beam splitter to a side collector or proceed to
the point of interference.  For a given interference result, we find
{}``population oscillations'' in the distributions of the deviated
particles, which are entirely controlled by the quantum angle.
Various versions of population oscillation experiments are discussed,
with two or three independent condensates.
\end{abstract}
\maketitle

\section{Introduction}

If two or more Bose-Einstein condensates (BEC) merge, they produce
a density interference pattern, as shown by spectacular experiments
with alkali atoms \cite{WK}. The usual explanation is that, when
spontaneous symmetry breaking (SSB) takes place at the Bose-Einstein
transition, each condensate acquires a random but well-defined phase.
The interference pattern then exhibits the relative phase. The simplest
form of this view involves the use of a classical complex variable
for each condensate given by

\begin{equation}
\left\langle \psi_{\alpha,\beta}(\mathbf{r})\right\rangle =\sqrt{n_{\alpha,\beta}(\mathbf{r})}e^{i\phi_{\alpha,\beta}(\mathbf{r)}}\label{eq:SSB}\end{equation}
 where $n_{\alpha,\beta}(\mathbf{r)}$ are the condensate densities
and $\phi_{\alpha,\beta}(\mathbf{r)}$ their phases. Another quantum
treatment of the problem can be carried out by the use of {}``phase
states,'' which describe a state of two condensates having a known
\emph{relative} phase and a fixed total number of particles \cite{Pethick}
- we will discuss the use of phase states in the next section. For
systems containing many particles the phase then appears as a macroscopic
quantity that has classical properties, but takes completely independent
random values from one realization of the experiment to the next.

\begin{figure}[h]
 \centering \includegraphics[width=3in]{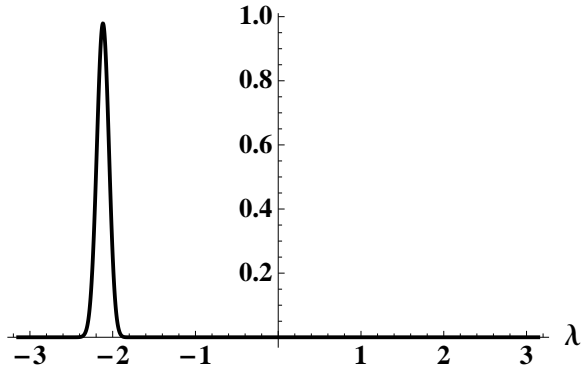}

\caption{The integrand $\prod_{i=1}^{M}\left(1+\cos(\mathbf{k}\cdot\mathbf{r}_{i}+\lambda)\right)$
of Eq.\ (\ref{classical}) as a function $\lambda$ after 200 measurements
at positions $\mathbf{r_{\mathit{\mathit{i}}}}.$ These first 200
measurements essentially convert a double Fock state into a state
resembling a phase state, peaked sharply at some particular value
$\lambda_{0},$, which is completely random from experiment to experiment.}

\label{fig1} 
\end{figure}

However, Bose-Einstein condensates are naturally described by Fock
states, states of definite particle number, for which the phase is
completely undetermined. Various authors \cite{Java,WCW,CGNZ,CD,M1,M2,FL,LM}
have shown that repeated quantum measurements of the relative phase
of two Fock states cause a well-defined value to emerge spontaneously,
but with a random value. The probability of finding $M$ particles,
out of a total of $N$, at positions $\mathbf{r}_{1},\cdots\mathbf{r}_{M}$
($M\ll N$) is shown to be given by

\begin{equation}
P(\mathbf{r}_{1},\cdots\mathbf{r}_{M})\sim\int_{-\pi}^{\pi}\frac{d\lambda}{2\pi}\prod_{i=1}^{M}\left[1+\cos(\mathbf{k}\cdot\mathbf{r}_{i}+\lambda)\right]\label{classical}\end{equation}
 where $\mathbf{k}$ is the wave number difference between the two
condensates. The product in the integrand can be interpreted as describing
the independent individual measurements of position with the interference
of two waves of relative phase $\lambda$, resulting in probability
$\left[(1+\cos(\mathbf{k}\cdot\mathbf{r}_{i}+\lambda)\right]/2$;
the $\lambda$ integration expresses that this phase is initially
completely unknown. Nevertheless, after a series of measurements has
been performed (still for $M\ll N$), the product of these probabilities
in Eq.\ (\ref{classical}) is found to peak sharply at some particular
value $\lambda_{0},$ which becomes better and better defined while
the experiments accumulate, but takes a completely uncorrelated random
value from experiment to experiment. Fig.\ \ref{fig1} illustrates
the peaking effect in the integrand in Eq.\ (\ref{classical}) after
200 measurements (the method by which we choose the position values
is given in Ref. \cite{MKL}).

Eq.\ (\ref{classical}) is quite capable of describing the interference
pattern seen in the MIT experiment \cite{MKL}. Note however that
the average over all possible phases makes the phase very similar
to the integrated variable $\lambda$ in Bell's theorem \cite{Bell},
or to an {}``element of reality''\ as defined by Einstein, Podolsky
and Rosen \cite{EPR} - and we know that this notion combined with
locality leads to contradictions with quantum mechanics. Eq.\ (\ref{classical})
can thus be seen as a {}``classical'' equation, which is unlikely
to be able to describe some truly quantum experiments (for instance,
it cannot violate Bell's theorem).

In some conditions, quantum interference effects arise so that the
description in terms of a classical phase is no longer sufficient;
a second angle (or its equivalent) becomes necessary: the {}``quantum
angle,'' which controls the amount of {}``quantumness'' in the
results of an interference experiment. This article discusses the
role of the quantum angle in general. While it is possible to carry
out such a discussion for the position measurements in free space,
as in Eq. (\ref{classical}), it turns out that interferometers with
dichotomic outputs provide especially interesting results, for instance
in terms of quantum non-locality; this is why interferometers will
be the central subject of this paper.

In Sec. \ref{Fockphase} we show how this quantum angle already appears
in a very simple situation, with one single beam splitter on which
two Fock states interfere; we relate the quantum angle to phase off-diagonal
terms. In Sec. \ref{Interferometeranalysis} we study the effects
of the quantum angle in an experiment with an interferometer providing
dicthotomic results in two different regions of space, and leading
to violations of the Bell inequalities. But other experiments involving
directly the quantum angle are also possible. One was suggested to
us by the recent article of Dunningham et al \cite{Dunn}, who considered
the interference pattern of three merging condensates and the resulting
{}``phase Schrödinger cat state'' formed by the remaining (non-measured)
particles. In Sec. \ref{Populationoscillations} we consider a simplified
version of this experiment with two condensates only, which interfere
on a beam splitter; among the total of $N$ particles, only $M$ interfere
and are detected at locations 1 and 2; the remaining are deflected
near their sources and separately counted in detectors 3 and 4 ($m_{\alpha}$
particles from condensate $\alpha$, and $m_{\beta}$ particles from
condensate $\beta$). For fixed numbers of such particles in detectors
1 and 2, the numbers found in detectors 3 and 4, as a function of
$m_{\alpha}$, are found to have an oscillating distribution - a {}``fringe''
pattern when plotted over an ensemble of such experiments. We will
see that this effect, which we call {}``population oscillations''
(PO), involves the interference of two peaks in the \emph{quantum
angle} distribution; thus such an experiment would also directly reveal
the existence of the quantum angle. One can show \cite{WMFLPRL} how
these oscillations represent an example of quantum interference of
macroscopically distinct states (QiMDS), a property of quantum mechanics
that can verify its validity in large scale systems \cite{Leggett}.

\section{A simple interferometer}

\label{Fockphase} In the derivation of \cite{LM,LM2,LM3}, both the
classical phase $\lambda$ and the quantum angle $\Lambda$ had similar
origins: conservation rules, which take the form of integrals over
these angles. Here we show that phase states can also be used to obtain
the same results, following a reasoning that is similar to that found,
for instance, in Ref.\ \cite{Pethick}. Mathematically, of course,
the two derivations are equivalent; but, physically, it is interesting
to obtain the same results from two different points of view.

\begin{figure}[h]
\centering \includegraphics[width=3in]{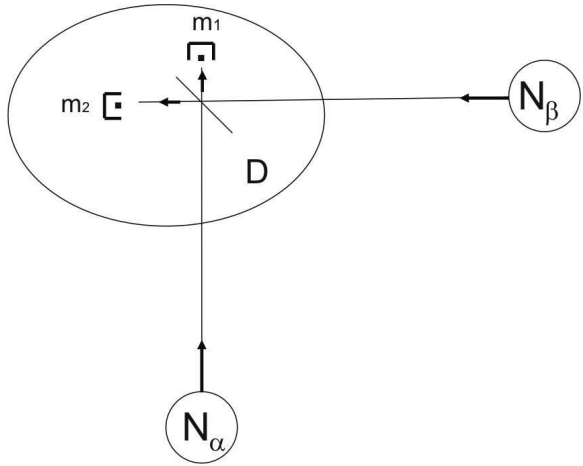}

\caption{Two Fock states, with populations $N_{\alpha}$ and $N_{\beta}$,
pass through a beam splitter, and are then made to interfere at detectors
1 and 2. }

\label{SimpleInterferometer} 
\end{figure}

We consider the experiment schematized in Fig. \ref{SimpleInterferometer},
where two Fock states with populations $N_{\alpha}$, and $N_{\beta}$
are emitted by two sources, cross a beam splitter, and interfere in
the regions of detection 1 and 2. Despite the apparent simplicity
of this device we have shown in a recent paper that remarkably complex
detector distributions can result \cite{SimpleInterferometer}. The
double Fock state describing the sources is 
\begin{equation}
\left|N_{\alpha},N_{\beta}\right\rangle =\frac{1}{\sqrt{N_{\alpha}!N_{\beta}!}}a_{\alpha}^{\dagger N_{\alpha}}a_{\beta}^{\dagger N_{\beta}}\left\vert \text{0}\right\rangle \label{initialstate}\end{equation}
where $\left\vert \text{0}\right\rangle$ is the vacuum state and  $a_{\alpha}^{\dagger}$ creates particle in state $\alpha$
corresponding to one source and $a_{\beta}^{\dagger}$ creates a $\beta$-state
particle corresponding to the other source. The total number of particles
is $N=N_{\alpha}+N_{\beta}$.

The destruction operators $a_{1}$ and $a_{2}$ associated with the
output modes can be written in terms of the mode operators at the
sources $a_{\alpha},$ $a_{\beta}$ by tracing back from the detectors
to the sources, with a phase shift of $\pi/2$ at each reflection:
\begin{equation}
a_{1}=\frac{1}{\sqrt{2}}\left[a+ia_{\beta}\right]\;;\; a_{2}=\frac{1}{\sqrt{2}}\left[ia_{\alpha}+a_{\beta}\right]\label{aa}\end{equation}

The probability amplitude describing the system after crossing the
beam splitter with $m_{1},m_{2}$ particles in the detector regions
is

\begin{equation}
C_{m_{1},m_{2}}=\left\langle
0\left|\frac{a_{1}^{m_{1}}a_{2}^{m_{2}}}{\sqrt{m_{1}!m_{2}!}}\right|N_{\alpha},N_{\beta}\right\rangle \label{C0}\end{equation}
 with $m_{1}+m_{2}=N$. To compute the state
 $a_{1}^{m_{1}}a_{2}^{m_{2}}\left|N_{\alpha},N_{\beta}\right\rangle $,
we expand the double Fock state in normalized (relative) phase states, defined for
two condensates (with constant total particle number $N$) as

\begin{eqnarray}
\left|\phi,N\right\rangle  & = & \frac{1}{\sqrt{2^{N}N!}}(a_{\alpha}^{\dagger}+e^{i\phi}a_{\beta}^{\dagger})^{N}\left|0\right\rangle =\nonumber \\
 & = & \frac{1}{\sqrt{2^{N}N!}}\sum_{n=0}^{N}\frac{N!}{\sqrt{n!(N-n)!}}\; e^{i\phi(N-n)}\left|n,N-n\right\rangle \label{eq:PhaseState}\end{eqnarray}
 where $N=N_{\alpha}+N_{\beta}$. The expansion in terms
of the phase states is \begin{equation}
\left|N_{\alpha}N_{\beta}\right\rangle =\sqrt{\frac{2^{N}N_{\alpha}!N_{\beta}!}{N!}}\int_{-\pi}^{\pi}\frac{d\phi}{2\pi}e^{-iN_{\beta}\phi}\left|\phi,N\right\rangle \label{Nphase}\end{equation}
 The action on phase states of the operators $a_{i}$ given in (\ref{aa})
is particularly simple; if we write them as 
\begin{equation}
a_{i}=v_{i\alpha}a_{i\alpha}+v_{i\beta}a_{\beta},
\end{equation}
 (with $v_{i\alpha}$ and $v_{i\beta}$ identified by Eqs. (\ref{aa})) we merely obtain%
\footnote{The phase state is obtained by repeated actions of the creation operator
$a_{\phi}^{\dagger}=(a_{\alpha}^{\dagger}+e^{i\phi}a_{\beta}^{\dagger})$
over vacuum, but no action of the {}``orthogonal'' creation operator
$a_{\phi+\pi/2}^{\dagger}$; the action of the annihilation operator
associated to the former operator is therefore simple, while that
of the latter gives zero. Expanding the $a_{i}$'s over the $a_{\alpha,\beta}$,
and then over $a_{\phi}$ and $a_{\phi+\pi/2}$, and keeping only
the component on the first annihilation operator, then directly leads
to (\ref{eq:aOperation}).%
} \cite{MKL}: \begin{equation}
a_{i}\left|\phi,N\right\rangle =\sqrt{\frac{N}{2}}(v_{i\alpha}+v_{i\beta}e^{i\phi})\left|\phi,N-1\right\rangle \label{eq:aOperation}\end{equation}
 Applying this result several times to (\ref{Nphase}) then gives
\begin{equation}
a_{1}^{m_{1}}a_{2}^{m_{2}}\left|N_{\alpha}N_{\beta}\right\rangle =\sqrt{\frac{N_{\alpha}!N_{\beta}!}{2^{N}}}\int_{-\pi}^{\pi}\frac{d\phi}{2\pi}R(\phi)\left|\,0\right\rangle \label{eq:PsiA}\end{equation}
 where $m_{1}+m_{2}=N$ and \begin{equation}
R(\phi)=2^{N/2}e^{-iN_{\beta}\phi}(v_{1\alpha}+v_{1\beta}e^{i\phi})^{m_{1}}(v_{2\alpha}+v_{2\beta}e^{i\phi})^{m_{2}}=e^{-iN_{\beta}\phi}(1+ie^{i\phi})^{m_{1}}(i+e^{i\phi})^{m_{2}}\label{eq:R}\end{equation}

When we insert this result into Eq.\ (\ref{C0}) and take the square
modulus, we obtain the probability in the form: \begin{equation}
P_{m_{1}m_{2}}=\frac{N_{\alpha}!N_{\beta}!}{m_{1}!m_{2}!}\int_{-\pi}^{\pi}\frac{d\phi^{\prime}}{2\pi}\int_{-\pi}^{\pi}\frac{d\phi}{2\pi}R^{*}(\phi^{\prime})R(\phi)\label{GEq}\end{equation}
 and find upon multiplying all these factors out:\begin{eqnarray}
P_{m_{1}m_{2}} & = & \frac{N_{\alpha}!N_{\beta}!}{m_{1}!m_{2}!}\int_{-\pi}^{\pi}\frac{d\phi^{\prime}}{2\pi}\int_{-\pi}^{\pi}\frac{d\phi}{2\pi}\cos[(N_{\alpha}-N_{\beta})\left(\phi-\phi^{\prime}\right)/2]\nonumber \\
 &  & \times\left[\cos\left(\frac{\phi-\phi^{\prime}}{2}\right)+\cos\left(\frac{\phi+\phi^{\prime}-\pi}{2}\right)\right]^{m_{1}}\left[\cos\left(\frac{\phi-\phi^{\prime}}{2}\right)-\cos\left(\frac{\phi+\phi^{\prime}-\pi}{2}\right)\right]^{m_{2}}\label{eq:GIntegrand}\end{eqnarray}
 It is then natural to make a variable change by introducing the average
of the two phases \begin{equation}
\lambda=(\phi+\phi^{\prime}-\pi)/2\end{equation}
 now identified as {}``the phase angle'', as well as the difference
\begin{equation}
\Lambda=(\phi-\phi^{\prime})/2\label{eq:QAdef}\end{equation}
 which we call the {}``quantum angle.'' Eq.\ (\ref{eq:GIntegrand})
then becomes

\begin{eqnarray}
P_{m_{1}m_{2}} & = & \frac{N_{\alpha}!N_{\beta}!}{N!}\frac{1}{m_{1}!m_{2}!}\int_{-\pi}^{\pi}\frac{d\Lambda}{2\pi}\int_{-\pi}^{\pi}\frac{d\lambda}{2\pi}\cos[(N_{\alpha}-N_{\beta})\Lambda]\nonumber \\
 &  & \times\left[\cos\left(\Lambda\right)+\cos\left(\lambda\right)\right]^{m_{1}}\left[\cos\left(\Lambda\right)-\cos\left(\lambda\right)\right]^{m_{2}}\label{eq:LamlamIntegrand}\end{eqnarray}
 This probability is a double sum over the variables $\lambda$ and
$\Lambda$ of a function of these variables as well as of the results
$m_{1}$ and $m_{2}$. According to (\ref{eq:QAdef}), if one sets
$\Lambda=0$ or $\Lambda=\pm\pi$ in this function, one obtains the
contributions of the terms that are diagonal in the phase representation.
The relevant values of the phase in the initial state then appear
directly. For instance, if the function has a single narrow peak around
some particular value, the phase is well-defined; if it has several
peaks at various values of the phase, for a pure state the system
is in a coherent superposition of different values of the phase (a
{}``Schrödinger cat'' if these values are very different and if
the system contains many particles). The role of the quantum angle
$\Lambda$ is precisely to signal the coherent character of the different
values of the phase (off-diagonal terms in the phase representation).
Each time non-zero values of this quantum angle play a role, the classical
description of Eq.\ (\ref{classical}) is not sufficient; the non-classical
behavior occurs because the factors $\left[\cos\left(\Lambda\right)\pm\cos\left(\lambda\right)\right]/2$
in the integrand of (\ref{eq:LamlamIntegrand}) can become negative,
so that they can no longer be interpreted as probabilities. In the ($\lambda$,$\Lambda$) plane, we will call the ``classical
region'' the region that lies around the $\lambda$ axis at $\Lambda=0$,
and the ``quantum region'' the rest of the plane.%
\footnote{The regions at $\Lambda=\pm\pi$ are also classical (i.e., equivalent
to $\Lambda=0$), as can be seen by showing that the integration segments
$-\pi\le\Lambda\le-\pi/2$ and $\pi/2\le\Lambda\le\pi$ (or equivalently
$\pi/2\le\Lambda\le3\pi/2$) give an identical contribution as the
region $-\pi/2\le\Lambda\le\pi/2$. To do so make the substitution
$\Lambda^{\prime}=\Lambda-\pi$ and $\lambda^{\prime}=\lambda-\pi$.}

\begin{figure}[h]
\includegraphics[width=3in]{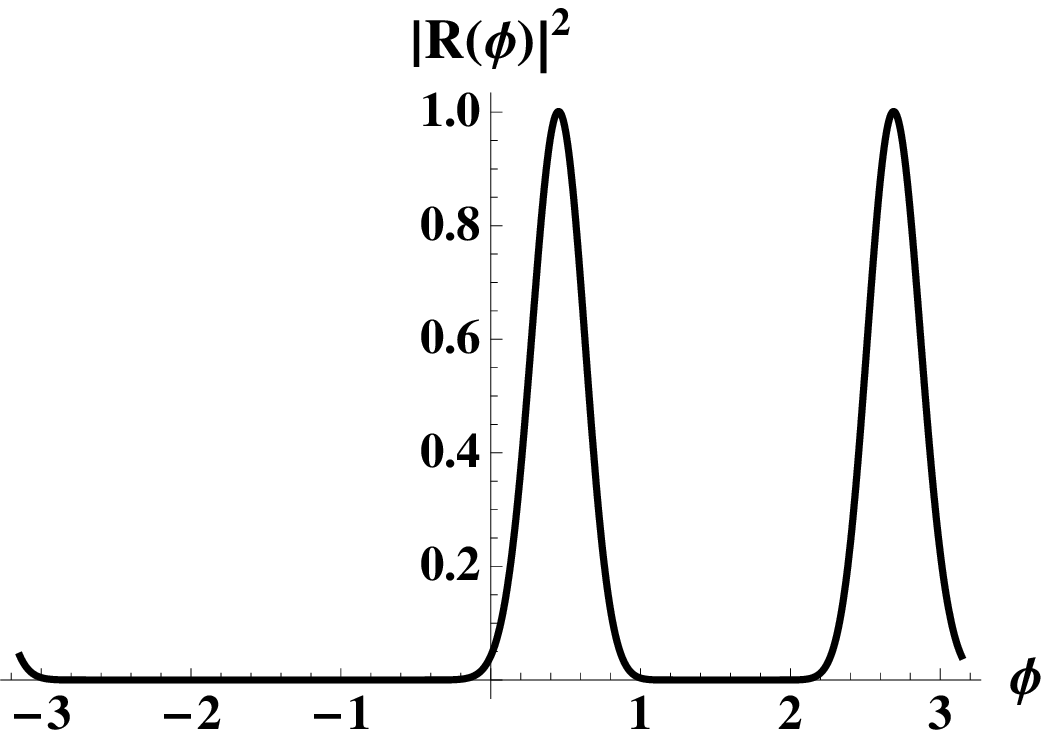}\quad{}\includegraphics[width=7cm]{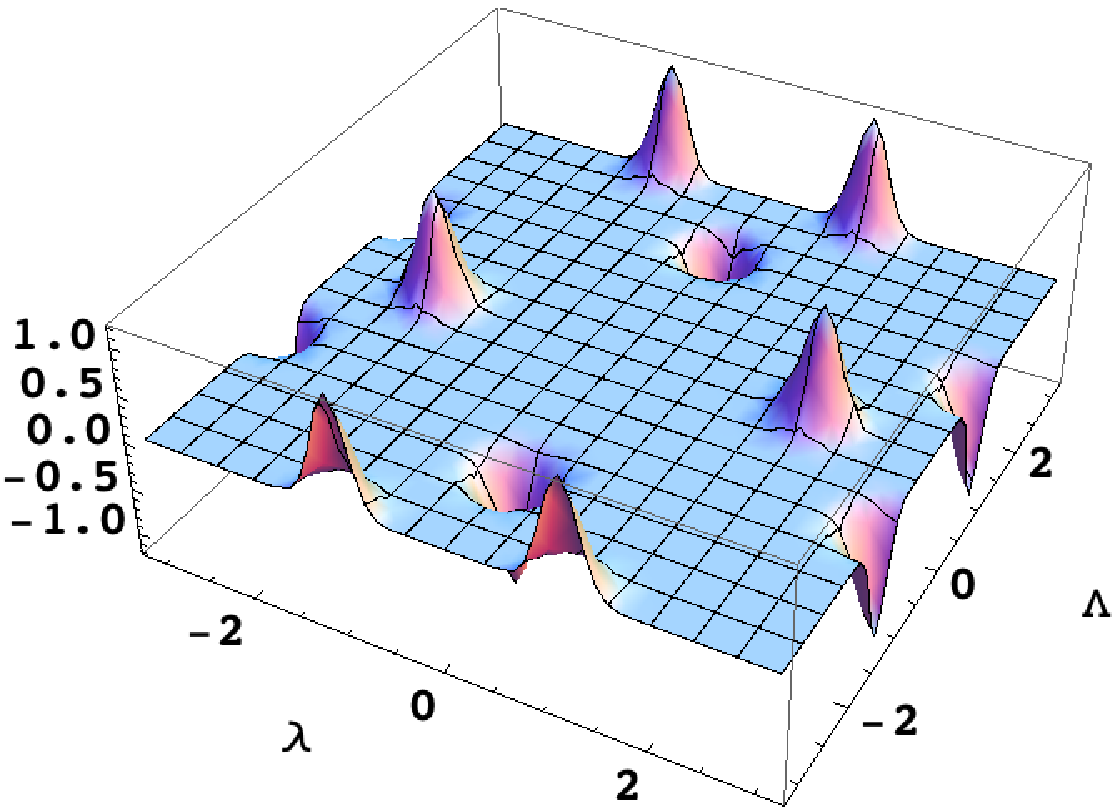}

\caption{(Color online) The left side (a) shows the coefficients $\left|R(\phi)\right|^{2}$
in an experiment where one finds $m_{1}=9$ and $m_{2}=23.$ The interferometer
is not able to distinguish two values of the phase difference $\phi_{0}$
between two sources; there remains an ambiguity between $\frac{\pi}{2}+\phi_{0}$
or $\frac{\pi}{2}-\phi_{0}$. Here $\phi_{0}=\pm1.12$. The right
side (b) shows the integrand of Eq.\ (\ref{eq:LamlamIntegrand})
for the same values of variables, and $N_{\alpha}=N_{\beta}$. Regions
on the lines $\Lambda=0,\pm\pi,$ correspond to diagonal terms in
relative phase, while those elsewhere represent off-diagonal contributions.}

\label{doublePeakinphi} 
\end{figure}

In Fig. \ref{doublePeakinphi}(a), we see the absolute square of the
coefficient $R$ of Eq.\ (\ref{eq:R}), showing two peaks for a particular
choice of $m_{1}$ and $m_{2}$ and with $N_{\alpha}=N_{\beta}$ and
$M=N$. This is not surprising since, classically, an ambiguity in
the sign of the phase angle difference also occurs in this interferometer:
two different values of this difference lead to the same intensities
in the two output arms. Fig. \ref{doublePeakinphi}(b) shows a plot
of the corresponding integrand of Eq.\ (\ref{eq:LamlamIntegrand}).
The diagonal phase contributions arise from the peaks on the lines
$\Lambda=0,\pm\pi$. Here the system is in a pure state, so that these
peaks are necessarily coherent; peaks in the quantum regions (away
from $\Lambda=0,\pm\pi.$) are also visible, which have a negative
sign and therefore signal destructive interference (for these particular
results of measurement; for other values, it is constructive).

\begin{figure}[h]
\centering \includegraphics[width=3in]{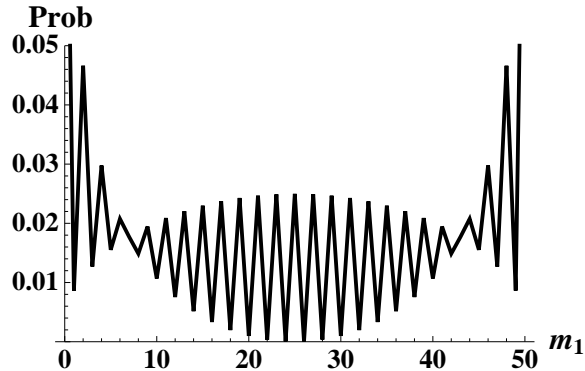}

\caption{The probability distribution of Eq. (\ref{eq:LamlamIntegrand}) for
input numbers $N_{\alpha}=26,$ $N_{\beta}=24.$}

\label{SimpleInterfPattern} 
\end{figure}

In Fig. \ref{SimpleInterfPattern} we show a particular example of
the probability distribution for finding the set of $\{m_{1},m_{2}\}$
particles in the detectors. The structure has a surprisingly complex
dependence on the numbers of particles in the Fock state inputs. The
simple interferometer is discussed more completely in a separate publication
\cite{SimpleInterferometer}. %

In this section we have recovered by the use of phase states the basic
results obtained from conservation rules in \cite{LM,LM2}. The present
method illustrates the relation between the two angles and the diagonal
or off-diagonal phase terms, and therefore the role of the classical
and non-classical region in the $\lambda$, $\Lambda$ plane. We now
examine how the quantum angle changes the description of some other
processes for Bose-Einstein condensates involving several interferometers.

\section{Double interferometer}

\label{Interferometeranalysis} 
\begin{figure}[h]
 . \centering \includegraphics[width=7cm]{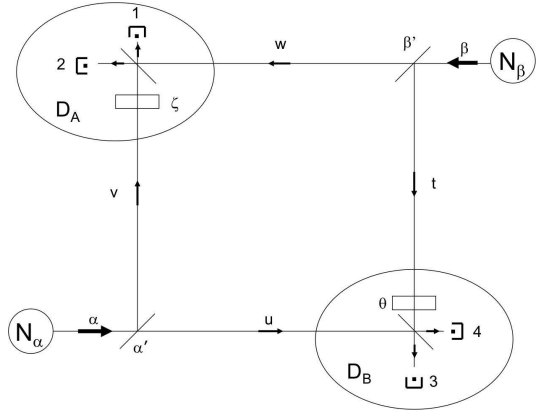}

\caption{Two Fock states, with populations $N_{\alpha}$ and $N_{\beta}$,
enter beam splitters, and are then made to interfere in two different
regions of space $D_{A}$ and $D_{B}$, with detectors 1 and 2 in
the former, 3 and 4 in the latter. In each of the channels $j=1,2,3,4$
particles are counted. We assume that no particle is missed: the sum
$M$ of the four $m_{j}$'s is equal to $N=N_{\alpha}+N_{\beta}$.}

\label{Interferometer} 
\end{figure}

We now discuss the role of the quantum
angle in an interferometer experiment designed to observe violations
of the Bell-Clauser-Horne-Shimony-Holt (BCHSH) inequality \cite{BCHSH},
already discussed in \cite{LM2}. The device is shown in Fig. \ref{Interferometer}
and involves a twin Fock state entering a double interferometer, which
can be used to measure the relative phase of the two condensates in
two remote regions of space. The relevance of twin Fock states for
phase measurements in simple interferometers was already discussed
in Ref.\ \cite{HollBur} in 1993. The measurement of the phase of
an arbitrary quantum state at different locations of space was discussed
in Ref.\ \cite{Franson} in 1994. A general discussion of the properties
of the quantum operator associated with the phase difference between
two modes can be found in Ref.\ \cite{BarPegg}. A more recent Ref.\ \cite{DragZin}
gives a discussion of the interference of two Fock states and of the
details of the statistics of the position measurements, in the context
of interferences in free space. %

\subsection{Quantum calculation}

For completeness, we briefly recall the quantum calculation in this
subsection. The destruction operators $a_{1}\cdots a_{4}$ associated
with the output modes can be written in terms of the mode operators
at the sources $a_{\alpha},a_{\beta},a_{\alpha^{\prime}}$ and $a_{\beta^{\prime}}$
by tracing back from the detectors to the sources, with a phase shift
of $\pi/2$ at each reflection, $\zeta$ or $\theta$ at the shifters,
and a $1/\sqrt{2}$ at each beam splitter. This gives the projections
of the two different source modes onto each detector mode \begin{eqnarray}
a_{1} & = & \frac{1}{2}\left[ie^{i\zeta}a_{\alpha}+ia_{\beta}\right];\qquad a_{2}=\frac{1}{2}\left[-e^{i\zeta}a_{\alpha}+a_{\beta}\right]\nonumber \\
a_{3} & = & \frac{1}{2}\left[ia_{\alpha}+ie^{i\theta}a_{\beta}\right];\qquad a_{4}=\frac{1}{2}\left[a_{\alpha}-e^{i\theta}a_{\beta}\right]\label{operators}\end{eqnarray}
 where we have eliminated $a_{\alpha^{\prime}}$ and $a_{\beta^{\prime}}$,
which contribute only vacuum. The source state, having $N_{\alpha}$
and $N_{\beta}$ particles in the two condensates is again given by
Eq.\ (\ref{initialstate}). \ The amplitude describing the system
crossing all beam splitters with $m_{1}\cdots m_{4}$ particles in
the detectors is

\begin{equation}
C_{m_{1},..,m_{4}}=\left\langle m_{1},m_{2},m_{3},m_{4}\right\vert \Phi\rangle~=\left\langle 0\right\vert \frac{a_{1}^{m_{1}}\cdots a_{4}^{m_{4}}}{\sqrt{m_{1}!\cdots m_{4}!}}\frac{a_{\alpha}^{\dagger N_{\alpha}}a_{\beta}^{\dagger N_{\beta}}}{\sqrt{N_{\alpha}!N_{\beta}!}}\left\vert 0\right\rangle \label{C}\end{equation}

The calculation is similar to that of Sec. \ref{Fockphase} and can
be found in Refs. \cite{LM2,LM3}. We substitute (\ref{operators})
into this expression, make binomial expansions of the sums, evaluate
the expectation value of the operators, replace Kronecker $\delta$'s
by integrals in the form $\delta_{N_{\gamma},p}=\int_{-\pi}^{\pi}\frac{d\lambda_{\gamma}}{2\pi}~e^{i(p-N_{\gamma})\lambda_{\gamma}}$
with $\gamma=\alpha,\beta$, and make an appropriate variable change.
We then obtain \begin{equation}
\mathcal{P}(m_{1},m_{2},m_{3},m_{4})=\frac{2^{-N}\, N_{\alpha}!N_{\beta}!}{m_{1}!\cdots m_{4}!}\int_{-\pi}^{\pi}\frac{d\lambda}{2\pi}\int_{-\pi}^{\pi}\frac{d\Lambda}{2\pi}\cos[(N_{\alpha}-N_{\beta})\Lambda]\prod_{i=1}^{4}\left[\cos\Lambda+\eta_{i}\cos\left(\lambda-\varphi_{i}\right)\right]^{m_{i}}\label{OldProb-first}\end{equation}
 where $\eta_{1}=\eta_{3}=1$; $\eta_{2}=\eta_{4}=-1$; $\varphi_{1}=\varphi_{2}=-\zeta$;
$\varphi_{3}=\varphi_{4}=\theta$. In \cite{LM3} we also consider
the case where only $M$ particles are measured among a total number
of $N$, by assuming losses of particles either near the sources or
near the detectors. The sum of the probabilities associated with the
orthogonal states corresponding to the result of measurement is then
\begin{equation}
\mathcal{P}(m_{1},m_{2},m_{3},m_{4})=\frac{2^{N-2M}M!(N/2!)^{2}}{N!\, m_{1}!\cdots m_{4}!}\int_{-\pi}^{\pi}\frac{d\lambda}{2\pi}\int_{-\pi}^{\pi}\frac{d\Lambda}{2\pi}[\cos\Lambda]^{N-M}\prod_{i=1}^{4}\left[\cos\Lambda+\eta_{i}\cos\left(\lambda-\varphi_{i}\right)\right]^{m_{i}}\label{OldProb}\end{equation}
 (for simplicity, from now on we assume that $N_{\alpha}=N_{\beta}$;
the probabilities have now been normalized to a total probability
of $1$ for all events associated with the detection of $M$ particles).

We have associated values of $\eta$ that are +1 for channels of detection
1 and 3, $-1$ for channels detectors 2 and 4. Assume now that Alice,
in the first detection region 1, calculates the product of all $\eta$
values that she obtains, that is the local parity $(-1)^{m_{2}}$,
which is called $\mathcal{A}=\pm1$; similarly Bob, in the second
detection region 2, calculates $\mathcal{B}=(-1)^{m_{4}}=\pm1$. We
then have two functions to which the BCHSH theorem can be applied.
The quantum average of their product is: \begin{equation}
\left\langle \mathcal{AB}\right\rangle =\sum_{m_{1}\cdots m_{4}}(-1)^{m_{2}+m_{4}}\mathcal{P}(m_{1},m_{2},m_{3},m_{4})\label{CAB}\end{equation}
 The result for the case where all particles are measured ($M=N)$
is found to be \cite{LM2}: \begin{equation}
\left\langle \mathcal{AB}\right\rangle =\left[\cos\left(\frac{\zeta+\theta}{2}\right)\right]^{N}\label{AB}\end{equation}

\subsection{Classical phase situations}

We consider the case where $M\ll N$ particles are detected; in (\ref{OldProb}),
the factor $\left[\cos\Lambda\right]^{N-M}$ is peaked sharply at
$\Lambda=0$.\ Setting $\cos\Lambda$ to unity in the product and
doing the integral over $\Lambda$ gives \begin{equation}
\mathcal{P}(m_{1},m_{2},m_{3},m_{4})=\frac{M!}{4^{M}m_{1}!\cdots m_{4}!}\int_{-\pi}^{\pi}\frac{d\lambda}{2\pi}\prod_{i=1}^{4}\left[1+\eta_{i}\cos\left(\lambda-\varphi_{i}\right)\right]^{m_{i}}\label{ClassP}\end{equation}
 where we have taken the $N\rightarrow\infty$ limit of the normalization
factor. The quantum angle $\Lambda$ has now disappeared from the
result, so that in the integrand all the terms in the product are
positive and can be interpreted as probabilities. The BCHSH inequality
\cite{BCHSH} then provides \begin{equation}
\left\langle \mathcal{AB}\right\rangle +\left\langle \mathcal{AB}^{\prime}\right\rangle +\left\langle \mathcal{A}^{\prime}\mathcal{B}\right\rangle -\left\langle \mathcal{A}^{\prime}\mathcal{B}^{\prime}\right\rangle \leq2\label{AB-bis}\end{equation}
 where letters with and without primes imply measurements at differing
angles. No violation of this inequality is possible as long as (\ref{ClassP})
applies.

This inequality can also be checked explicitly by computing the value
of the average $\left\langle \mathcal{AB}\right\rangle $ from (\ref{ClassP});
we find \begin{equation}
\left\langle \mathcal{AB}\right\rangle =\frac{M!}{(\frac{M}{2}!)^{2}2^{M}}\left[\cos\left(\frac{\zeta+\theta}{2}\right)\right]^{M}\label{ABeq}\end{equation}

\subsection{Fully quantum situations}

We now assume that all particles are measured. For convenience, Alice's
measurement angle is taken as $\phi_{a}=\zeta/2$ and Bob's as $\phi_{b}=-\theta/2$.
We define $E(\phi_{a}-\phi_{b})=\cos^{N}\left(\phi_{a}-\phi_{b}\right)$,
set $\phi_{a}-\phi_{b}=\phi_{b}-\phi_{a^{\prime}}=\phi_{b^{\prime}}-\phi_{a}=\xi$
and $\phi_{b^{\prime}}-\phi_{a^{\prime}}=3\xi$. We now maximize $Q=3E(\xi)-E(3\xi)$
in order to find the greatest violation of the inequality for each
$N$. For $N=2$ we find $Q_{\max}=2.41$ at $\xi=0.39$; for $N=4,$
$Q_{\max}=2.36$ at $\xi=0.26$; and for $N\rightarrow\infty,$ $Q_{\max}\rightarrow2.32$
with $\xi\cong0.52/\sqrt{N}$. The system continues to violate local
realism for arbitrarily large condensates.

Despite the identical dependence in the cosine factor in (\ref{ABeq})
and (\ref{AB}), the effect of the prefactor, always equal to or less
than 1/2 in the classical case, is to prevent the violations of the
inequalities to occur. Actually, quantum violations disappear even
when only \emph{one} particle is missed in the measurement process
($M=N-1$) as discussed in Ref. \cite{LM2}.

\subsection{Discussion}

\begin{figure}[t]
 \includegraphics[width=3in]{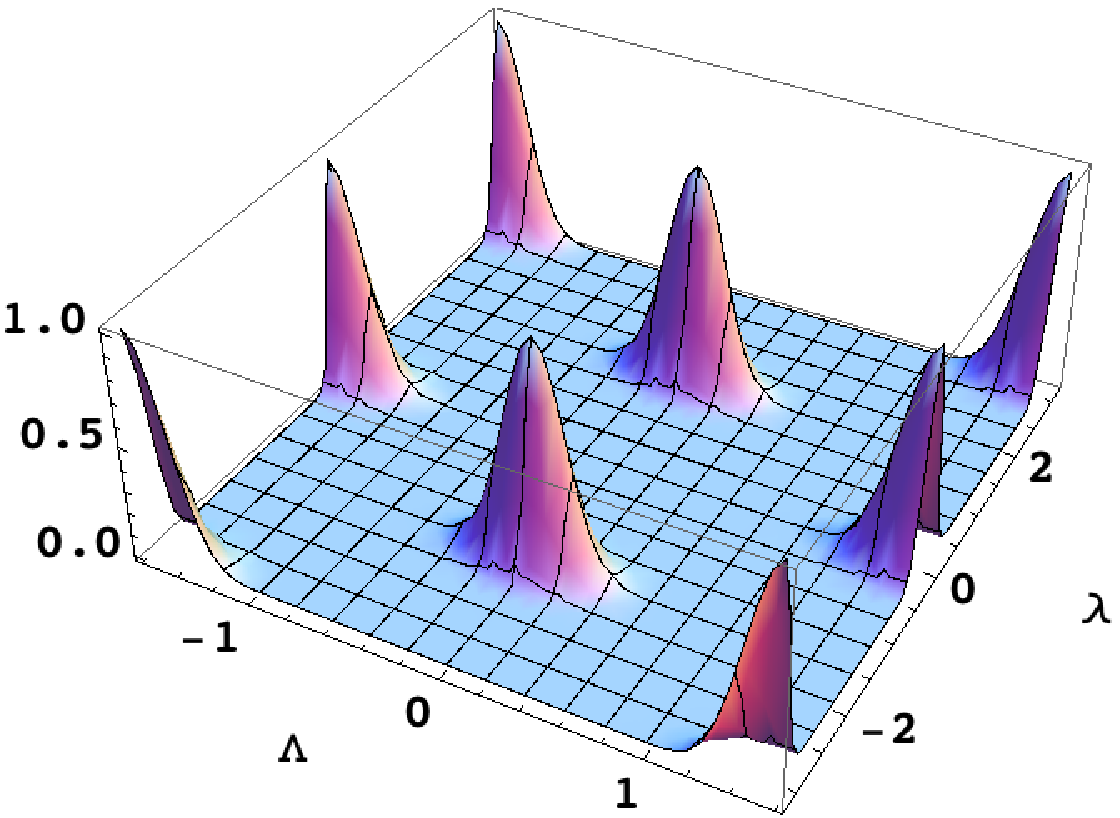}
\hspace{0.2in} \includegraphics[width=3in]{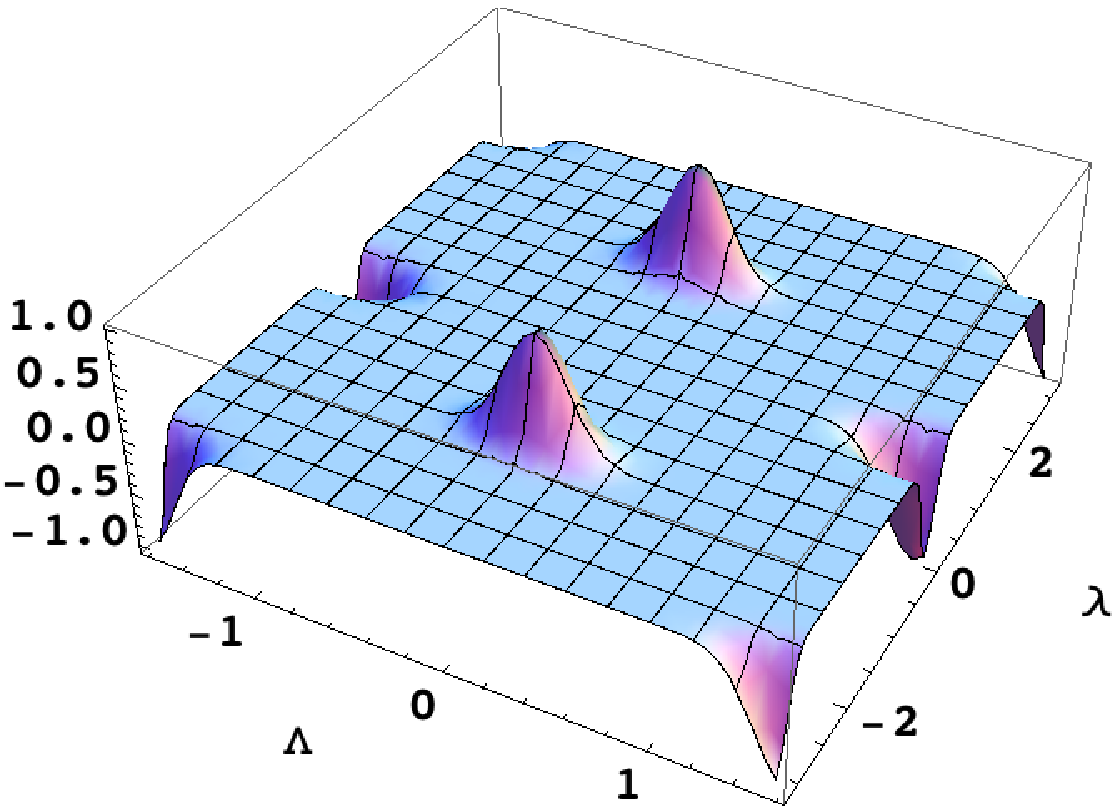}
\caption{(Color online) Plot as a function of $\lambda$ and $\Lambda$ of the integrand
in Eq.\ (\ref{OldProb}), for $N_{\alpha}=N_{\beta}=M=40$, and $\theta=\zeta=0$.
Left: $m_{2}+m_{4}=20$; right $m_{2}+m_{4}=19$. Depending on the
parity of this sum, the peaks in the {}``quantum region'' $\Lambda\ne0$
have the opposite sign; this indicates that the measurements of parities
should be a good choice of local observables to obtain strong violations
of the BCHSH inequalities}

. \label{Interferometer2} 
\end{figure}

It is interesting to see in more detail how the quantum angle is involved
in the BCHSH violation. For instance, Fig. \ref{Interferometer2}
shows the variations as a function of $\lambda$ and $\Lambda$ of
the function that appears in the integral of Eq. (\ref{OldProb}),
for $N_{\alpha}=N_{\beta}=M=40$, and $\theta=\zeta=0$. The left
part of the figure assumes that $m_{1}=6$, $m_{2}=14$, $m_{3}=14$
and $m_{4}=6$, the right part that $m_{1}=6$, $m_{2}=14$, $m_{3}=15$
and $m_{4}=5$; one immediately notices that, depending on the parity
of the sum $m_{2}+m_{4}$, the peaks in the {}``quantum region''
$\Lambda\ne0$ have the opposite sign. This explains why the quantum
effects will be enhanced if Alice and Bob decide to choose the parities
(product of all their results $\eta$'s) as their local observables
$\mathcal{A}$ and $\mathcal{B}$. It is then natural that strong
violations of the BCHSH inequalities should be obtained for this particular
choice, while of course Alice and Bob could combine their local results
in many other ways to obtain functions $\mathcal{A}$ and $\mathcal{B}$.

Suppose now we delete the leading normalization factors in each of
Eqs.\ (\ref{OldProb}) and (\ref{ClassP}) and then evaluate the
unnormalized values of $\left\langle \mathcal{AB}\right\rangle $
for $M=2.$ The result in each case is $4$$\left[\cos\left(\frac{\zeta+\theta}{2}\right)\right]^{2}$.
Thus the entire difference between quantum and classical averages
is in the normalization given, respectively, by the integrals over
$\Lambda$ and $\lambda$ of \begin{eqnarray}
L_{qu}(\xi,\Lambda,\lambda) & = & \sum_{m_{1}\cdots m_{4}}^{\prime}\left[\cos\Lambda+\eta_{i}\cos\left(\lambda+\zeta\right)\right]^{m_{1}}\left[\cos\Lambda+\eta_{i}\cos\left(\lambda+\zeta\right)\right]^{m_{2}}\nonumber \\
 &  & \times\left[\cos\Lambda+\eta_{i}\cos\left(\lambda-\theta\right)\right]^{m_{3}}\left[\cos\Lambda+\eta_{i}\cos\left(\lambda-\theta\right)\right]^{m_{4}}\end{eqnarray}
 where the sum is on all $m_{i}$ totaling 2; and\begin{eqnarray}
L_{cl}(\xi,\lambda) & = & \sum_{m_{1}\cdots m_{4}}^{\prime}\left[1+\eta_{i}\cos\left(\lambda+\zeta\right)\right]^{m_{1}}\left[1+\eta_{i}\cos\left(\lambda+\zeta\right)\right]^{m_{2}}\nonumber \\
 &  & \times\left[1+\eta_{i}\cos\left(\lambda-\theta\right)\right]^{m_{3}}\left[1+\eta_{i}\cos\left(\lambda-\theta\right)\right]^{m_{4}}\end{eqnarray}
 For $M=2$ we explicitly get\begin{eqnarray}
L_{qu}(\xi,\Lambda,\lambda) & = & 8\cos^{2}\Lambda\\
L_{cl}(\xi,\lambda) & = & 8\end{eqnarray}
 The quantum normalization integrand clearly yields a smaller normalization
integral enhancing the $\left\langle \mathcal{AB}\right\rangle $
average and allowing the violation of the BCHSH inequality. It is
this variation with quantum angle that allows the violation.

\section{Population oscillations}

\label{Populationoscillations} Dunningham \emph{et al} \cite{Dunn}
have considered a situation in which three condensates, a, b, and
c, each contain initially $N/3$ particles. A number of them, $M<N$,
form an interference pattern on a screen, while the remaining particles
$m_{a},m_{b,}$ and $m_{c}$ are counted elsewhere (perhaps having
been deflected by beam splitters while traveling from the sources),
or in a second step of the experiment. The numbers of such particles,
as a function of $m_{a}$ and $m_{b}$, are found to have an oscillating
distribution when plotted over an ensemble of experiments corresponding
to the same interference pattern for the first $M$ particles. This
phenomenon was explained as arising from the interference of the two
coherent components of a phase {}``Schrödinger cat state'' of the
system.

\subsection{Population oscillations by two-source interferometer}

Here we present a simpler version of this experiment based on the
interferometer shown in Fig. \ref{PO interferometer}, which nevertheless
retains the essential features of the three condensate device. The
general idea is that condensates provide, in a sense, many realizations
of the same single particle quantum state, since they contain many
particles in the same individual state. One can then perform experiments
where some particles are used to measure one quantum observable, some
others another {}``incompatible'' observable, which would be impossible
with one single realization of the quantum state. In this case, the
incompatible (non-commuting) observables will be the phase and the
number of particles.

\subsubsection{Experimental setup}

In our version of the experiment, $M$ particles from the two sources
interfere in the detector D made up of a beam splitter and subdetectors
1 and 2; the other particles are detected before they reach the interferometer,
with the help of additional beam splitters followed by detectors 3
and 4. For the sake of simplicity, we assume that all beam splitters
have 1/2 reflectivity and transmissivity; we write $m_{1}$ and $m_{2}$
as the number of particles seen at subdetectors 1 and 2, respectively
(with $M=m_{1}+m_{2}$), $m_{\alpha}$ and $m_{\beta}$ the number
of particles seen in detectors 3 and 4. In this scheme, some of the
particles are used to measure the relative phase of the two sources
$N_{\alpha}$ and $N_{\beta}$, the others to obtain information about
their initial populations.

\begin{figure}[h]
\centering \includegraphics[width=7cm]{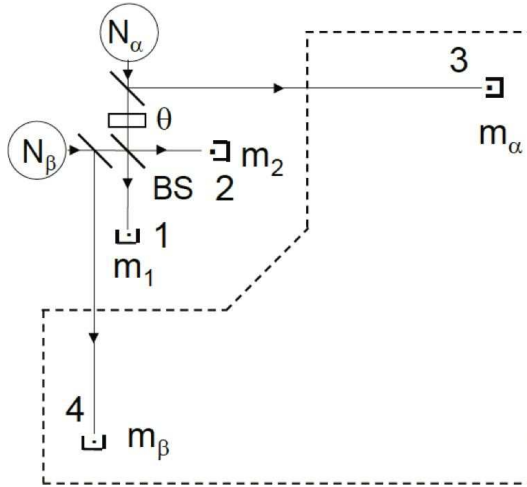} \caption{Two source condensates states, with populations $N_{\alpha}$ and
$N_{\beta}$, emit particles that cross beam splitters. Some particles
reach the central beam splitter followed by detectors 1 and 2, registering
$m_{1}$ and $m_{2}$ counts. The other particles are then described
by a quantum superposition of macroscopically distinct states propagating
inside the region shown with a dotted line; they eventually reach
counters 3 and 4, which register $m_{\alpha}$ and $m_{\beta}$ counts
respectively. A phase shift \textgreek{j} = \textgreek{p}/2 occurs
in one path.}

\label{PO interferometer} 
\end{figure}

Assume for a moment that the central beam splitter is removed, so
that no interference effect between the sources takes place at detectors
1 and 2. Then the experiment separates into two independent parts:
the detectors 1 and 3 measure the population of one source, and the
sum $m_{1}+m_{\alpha}$ gives an exact measurement of the initial
population $N_{\alpha}$; of course $m_{1}$ and $m_{\alpha}$ may
fluctuate separately, with a constant sum, but their most likely value
is $N_{\alpha}/2$. Similarly, detectors 2 and 4 give information
on the population of the other source, and the most likely number
of their counts is $N_{\beta}/2$.

Now, when the central beam splitter is inserted, the counts of detectors
1 and 2 can no longer be ascribed to any of the sources, which are
indistinguishable for the detectors; what they actually measure is
their relative phase. In classical optics for instance, if the sources
are lasers with the same intensity and a phase difference $\phi_{\alpha\beta}$,
the numbers of counts $m_{1}$ and $m_{2}$ are respectively proportional
to $\sin^{2}(\lambda/2)$ and $\cos^{2}(\lambda/2)$, with %
\footnote{A phase shift $\pi/2$ is introduced by each reflection on a beam
splitters%
} $\lambda=\phi_{\alpha\beta}-\pi/2$; the counting rates therefore
provide information about the absolute value $\lambda$, but not its
sign. In quantum mechanics, this sign uncertainty becomes an essential
ingredient for the creation of a superposition of two states with
different phases (a {}``Schrödinger cat''): the measurement process
at the interferometer projects the initial state of the system onto
two categories of phase states with opposite phase difference, between
which no selection is made. Therefore, after the interference measurement,
the system reaches a coherent superposition of states with opposite
values of the phase.

How can this superposition be observed? The conjugate variable of
the relative phase is the population difference between the sources;
therefore, as the authors of Ref.\ \cite{Dunn} have remarked, if
one measures the absolute value of this difference, one expects to
see interference effects between the two components of the coherent
state with opposite signs for the phase. Fortunately, even with the
central beam splitter inserted, detectors 3 and 4 can still be used
to obtain information about the populations of the sources. So, for
one given value of the ratio $m_{1}/m_{2}$, one expects oscillations
of the probabilities associated with given values of $m_{3}$ and
$m_{4}$, that is {}``population oscillations''. This is the general
physical idea, based on the fact that Fock states provide many realizations
of one single particle quantum state, as mentioned in the introduction
of this section. We will see that, in our analysis, the interference
that produces the oscillations occurs between peaks in the quantum-angle
distribution.

Leggett \cite{Leggett} has considered how one might observe coherent
superpositions of large numbers of particles by observing their interferenece
({}``quantum interference of macroscopically distinct states'' or
QIMDS). One can tell the difference between such a pure state and
a statistical mixture only by observing the off-diagonal matrix elements
between the different wave function elements. Our population oscillations
are the result of such an interference as we will discuss below. 

The experimental setup of Fig. \ref{PO interferometer} is completely
defined, as required in the Copenhagen view of quantum mechanics;
in particular, the setup does not have to be changed from an interference
setup to a population measurement setup in the middle of the experiment.
We now calculate the probabilities associated with the various possible
results of measurements.

\subsubsection{Qualitative analysis}

\begin{figure}[h]
\centering \includegraphics[width=7cm]{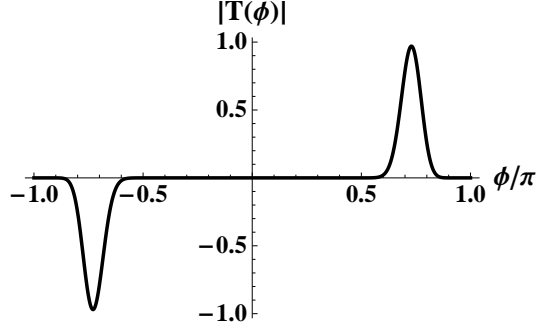}
\caption{Variations of $\hat{R}(\phi)$ obtained for $m_{1}=17$ and $m_{2}=83$.
The peaks are at $\phi_{0}=\pm0.73\pi$ (the phase choice $\theta=\pi/2$
gives symmetrical peaks about zero). The relative sign of the two
peaks is (-1)$^{m_{2}}$. For large numbers of particles, the measurement
produces a coherent superposition of macroscopically distinct states
(\textquotedblleft{}Schrödinger cat\textquotedblright{}).}

\label{CatPicture} 
\end{figure}

We assume that all particles are detected; the total number then is
$N=N_{\alpha}+N_{\beta}=m_{1}+m_{2}+m_{\alpha}+m_{\beta}$. We will
vary the number of particles in detectors 3 and 4 at constant $N,M,m_{1},m_{2}$
to examine the behavior of the probability on the set $\{m_{\alpha,}m_{\beta}\}$.
The destruction operators for particles at the detectors in terms
of the source destruction operators are\begin{eqnarray}
a_{1} & = & \frac{1}{2}\left(a_{\alpha}+ia_{\beta}\right);\qquad a_{2}=\frac{1}{2}\left(ia_{\alpha}+a_{\beta}\right)\nonumber \\
a_{3} & = & \frac{1}{\sqrt{2}}a_{\alpha};\qquad a_{4}=\frac{1}{\sqrt{2}}a_{\beta}\end{eqnarray}
 The probability amplitude for detecting the set $\{m_{1},m_{2},m_{\alpha}m_{\beta}\}$
is given by\begin{equation}
C_{m_{1}m_{2},m_{\alpha},m_{\beta}}=\frac{1}{\sqrt{m_{1}!m_{2}m_{\alpha}!m_{\beta}!}}\left\langle 0\right|a_{1}^{m_{1}}a_{2}^{m_{2}}a_{3}^{m_{\alpha}}a_{4}^{m_{\beta}}\left|N_{\alpha}N_{\beta}\right\rangle \end{equation}
Expand the double Fock state in phase states (Eq. (\ref{Nphase}))
and operate with $a_{1}^{m_{1}}a_{2}^{m_{2}}$ so the state created
by the interferometer detectors 1 and 2 is \begin{equation}
\left|\Gamma\right\rangle \equiv a_{1}^{m_{1}}a_{2}^{m_{2}}\left|N_{\alpha}N_{\beta}\right\rangle =\sqrt{\frac{N_{\alpha}!N_{\beta}!}{2^{N}}}\int_{-\pi}^{\pi}\frac{d\phi}{2\pi}e^{-iN_{\beta}\phi}R(\phi)\left|\phi,N-M\right\rangle \label{eq:Gamma}\end{equation}
 where $m_{1}+m_{2}=M$ and \begin{equation}
R(\phi)=(e^{i\theta}+ie^{i\phi})^{m_{1}}(ie^{i\theta}+e^{i\phi})^{m_{2}}\end{equation}
If we take $\theta=\pi/2$ then $R(\phi)$ takes the simple form\begin{equation}
R(\phi)=(2ie^{i\phi/2})^{M}\left(\cos\frac{\phi}{2}\right)^{m_{1}}\left(\sin\frac{\phi}{2}\right)^{m_{2}}\end{equation}
Figure \ref{CatPicture} shows ${T(\phi)}=R(\phi)(2ie^{i\phi/2})^{-M}$,
which has two peaks at $\pm\phi_{0}=\pm\arctan\sqrt{m_{2}/m_{1}}$.
This is not surprising since, classically, the ratios of the intensities
in the output arms of the interferometer determines the absolute value
of the phase difference between the two input arms but not its sign.
Separating negative and positive contributions of $\phi$ provides\begin{equation}
\left|\Gamma\right\rangle =\left|\psi_{+}\right\rangle +(-1)^{m_{2}}\left|\psi_{-}\right\rangle \end{equation}
where \begin{equation}
\left|\psi_{\pm}\right\rangle \sim e^{\mp i(N_{\beta}-M/2)\phi_{0}}\left|\pm\phi_{0},N-M\right\rangle \end{equation}
We assume that $M$ is large, so that the peaks are sharp and these
two branches are orthogonal for any $\phi_{0}$ not too near zero;
and they are macroscopic as long as $N-M$ is large. The interference
between these two states (QIMDS) is provided by the side detectors
in Fig. \ref{PO interferometer}. Because we have \begin{equation}
a_{1}^{m_{1}}a_{2}^{m_{2}}\left|\pm\phi_{0},N-M\right\rangle \sim e^{\pm im_{\beta}\phi_{0}}\left|0\right\rangle \end{equation}
then the probability of gettting the set $\{m_{1},m_{2},m_{\alpha},m_{\beta}\}$
is \begin{equation}
P(m_{1},m_{2},m_{\alpha},m_{\beta})\sim1+(-1)^{m_{2}}\cos[(m_{\alpha}-m_{\beta})\phi_{0}]\label{eq:QualitativePOForm}\end{equation}
where we have taken $N_{\alpha}=N_{\beta}.$ The cosine terms in this
come from the two cross terms $\left\langle \psi_{\pm}\left|a_{3}^{\dagger m_{\alpha}}a_{4}^{\dagger m_{\beta}}a_{3}^{m_{\alpha}}a_{4}^{m_{\beta}}\right|\psi_{\mp}\right\rangle .$
If one does the interferometer experiment for fixed source numbers,
say, $N_{\alpha}=N_{\beta}$, and considers only those experiments
having the same $m_{1},m_{2}$ then the interference between the two
elements will show up in a cosine variation of probability with $m_{\alpha}$.
We call this effect \textquotedblleft{}population oscillations.\textquotedblright{}
These oscillations are beyond SSB since they disappear if one starts
from either of Eqs. (\ref{eq:SSB}) or (\ref{eq:PhaseState}). With
a phase state of phase $\chi$ for instance, the action of the destruction
operators $a_{1,2}$ on this state introduces $\chi$ instead of an
integration variable $\phi$ into Eq. (\ref{eq:Gamma}) without the
$\phi$ integral. No interference effect between two phase peaks occurs
and the probability is proportional to $|R(\chi)|^{2}$. One gets
a $m_{\alpha},m_{\beta}$ dependence of the probability that is proportional
to a simple binomial distribution $(N-M)!/m_{\alpha}!m_{\beta}!$,
without any oscillation. Actually the angle $\chi$ plays no role
at all in this dependence, which is natural since detectors 3 and
4 do not see an interference effect between two beams; they just measure
the intensities of two independent sources after a beam splitter at
their output.

\subsubsection{Exact Quantum calculation}

The probability amplitude for detecting the set $\{m_{1},m_{2},m_{\alpha}m_{\beta}\}$
can be manipulated differently:\begin{eqnarray}
C_{m_{1}m_{2},m_{\alpha},m_{\beta}} & = & \frac{1}{\sqrt{m_{1}!m_{2}m_{\alpha}!m_{\beta}!N_{\alpha}!N_{\beta}!}}\left\langle 0\right|a_{1}^{m_{1}}a_{2}^{m_{2}}a_{3}^{m_{\alpha}}a_{4}^{m_{\beta}}a_{\alpha}^{\dagger N_{\alpha}}a_{\beta}^{\dagger N_{\beta}}\left|0\right\rangle \nonumber \\
 & = & \frac{\sqrt{N_{\alpha}!N_{\beta}!}}{\sqrt{m_{1}!m_{2}!m_{\alpha}!m_{\beta}!}}\frac{1}{\left(\sqrt{2}\right)^{m_{\alpha}+m_{\beta}}2^{m_{1}+m_{2}}}\sum_{p,q}\frac{m_{1}!}{p!(m_{1}-p)!}\frac{m_{2}!}{q!(m_{2}-q)!}\nonumber \\
 &  & \times i^{m_{1}-p}i^{q}\delta_{p+q+m_{\alpha},N_{\alpha}}\delta_{m_{1}+m_{2}-p-q+m_{\beta}.N_{\beta}}\nonumber \\
 & = & \sqrt{\frac{m_{1}!m_{2}!N_{\alpha}!N_{\beta}!}{m_{\alpha}!m_{\beta}!}}\frac{i^{N_{\alpha}+m_{1}-m_{\alpha}}}{\left(\sqrt{2}\right)^{m_{\alpha}+m_{\beta}}2^{m_{1}+m_{2}}}\nonumber \\
 &  & \times\sum_{p=0}^{m_{1}}\frac{(-1)^{p}}{p!(m_{1}-p)!(N_{\alpha}-m_{\alpha}-p)!(p+m_{\alpha}+m_{2}-N_{\alpha})!}\label{Amp2Cond}\end{eqnarray}

The probability of getting the set $\{m_{1,}m_{2},m_{\alpha},m_{\beta}\}$
for the sources numbers $N_{\alpha},N_{\beta}$ is then \begin{eqnarray}
P(m_{1},m_{2},m_{\alpha},m_{\beta}) & = & \frac{m_{1}!m_{2}!N_{\alpha}!N_{\beta}!}{m_{\alpha}!m_{\beta}!2^{m_{1}+m_{2}}2^{N}}\nonumber \\
 &  & \times\left[\sum_{p=0}^{m_{1}}\frac{(-1)^{p}}{p!(m_{1}-p)!(N_{\alpha}-m_{\alpha}-p)!(p+m_{\alpha}+m_{2}-N_{\alpha})!}\right]^{2}\label{Prob2Cond}\end{eqnarray}
a result that allows simple numerical computations. 

An alternative form suitable for illustrating the phase relations
is obtained if we choose to replace one of the $\delta$-functions
in Eq.\ (\ref{Amp2Cond}) by an integral, that is\[
\delta_{p+q+m_{\alpha},N_{\alpha}}=\int_{-\pi}^{\pi}\frac{d\phi}{2\pi}e^{i(p+q+m_{\alpha}-N_{\alpha})\phi}.\]
 The other $\delta$-function simply requires $N=m_{1}+m_{2}+m_{\alpha}+m_{\beta}.$
For the amplitude we then get \begin{equation}
C_{m_{1}m_{2},m_{\alpha},m_{\beta}}=\frac{\sqrt{N_{\alpha}!N_{\beta}!}}{\sqrt{m_{1}!m_{2}!m_{\alpha}!m_{\beta}!}}\frac{1}{\left(\sqrt{2}\right)^{m_{\alpha}+m_{\beta}}2^{m_{1}+m_{2}}}\int_{-\pi}^{\pi}\frac{d\phi}{2\pi}e^{-i(N_{\alpha}-m_{\alpha})\phi}(e^{i\phi}+i)^{m_{1}}(ie^{i\phi}+1)^{m_{2}}\end{equation}
 Squaring $C$ introduces another angle $\phi^{\prime}.$ A change
of variables to the relative phase angle \begin{equation}
\lambda=(\phi+\phi^{\prime}-\pi)/2\end{equation}
 and the quantum angle \begin{equation}
\Lambda=(\phi-\phi^{\prime})/2\end{equation}
 gives the form \begin{eqnarray}
P(m_{1},m_{2},m_{\alpha},m_{\beta}) & = & \frac{N_{\alpha}!N_{\beta}!}{m_{1}!m_{2}!m_{\alpha}!m_{\beta}!2^{N}}\int_{-\pi}^{\pi}\frac{d\lambda}{2\pi}\int_{-\pi}^{\pi}\frac{d\Lambda}{2\pi}e^{-i\left(N_{\alpha}-m_{\alpha}-N_{\beta}+m_{\beta}\right)\Lambda}\nonumber \\
 &  & \times\left[\cos\Lambda+\cos\lambda\right]^{m_{1}}\left[\cos\Lambda-\cos\lambda\right]^{m_{2}}\nonumber \\
 & = & \frac{N_{\alpha}!N_{\beta}!}{m_{1}!m_{2}!m_{\alpha}!m_{\beta}!2^{N}}\int_{-\pi}^{\pi}\frac{d\lambda}{2\pi}\int_{-\pi}^{\pi}\frac{d\Lambda}{2\pi}\cos\left[\left(N_{\alpha}-m_{\alpha}-N_{\beta}+m_{\beta}\right)\Lambda\right]\nonumber \\
 &  & \times\left[\cos\Lambda+\cos\lambda\right]^{m_{1}}\left[\cos\Lambda-\cos\lambda\right]^{m_{2}}\label{eq4}\end{eqnarray}
 Again we see the appearance of the quantum angle $\Lambda.$ We can
limit the integration over $\Lambda$ to non-redundant regions by
noting that a segment from $\pi/2$ to $3\pi/2$ is identical to that
from just $-\pi/2$ to $\pi/2$, as seen by making the substitutions
$\Lambda^{\prime}=\Lambda-\pi$ and $\lambda^{\prime}=\lambda-\pi$. (Cf.
footnote 2 in Sec. II.)

\begin{figure}[h]
\centering \includegraphics[width=7cm]{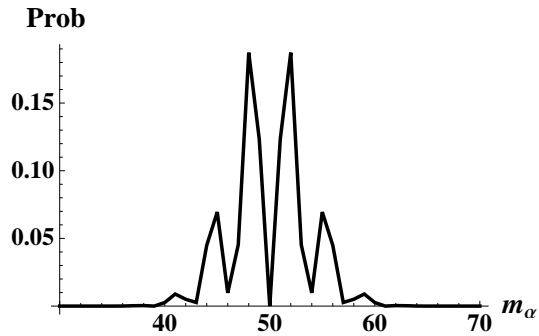}
\caption{Plot of $P(m_{1},m_{\alpha})$ given by Eq. (\ref{Prob2Cond}) versus
$m_{\alpha}$ for $N_{\alpha}=N_{\beta}=M=100,$ $m_{1}=17$ and $m_{2}=83$.
If $m_{2}$ is even, the central dip is replaced by a peak.}

\label{PO1783} 
\end{figure}

A typical example of a population oscillation plot computed from Eq.
(\ref{Prob2Cond}) is shown in Fig.  \ref{PO1783}.  We use $N_{\alpha}
= N_{\beta}$, but from Eqs (\ref{Prob2Cond}) and (\ref{eq4}) we see that the result would
not change if we took $N_{\alpha} \neq N_{\beta}$.

\subsubsection{Classical and quantum regions for the distribution}

In Eq.\ (\ref{eq4}) the $m_{\alpha}$ and $m_{\beta}$ dependencies
appear as a cosine Fourier transform with respect to the quantum angle
$\Lambda$ variable; this cosine Fourier transform is therefore the
origin of the population oscillations. If $\Lambda$ is set to zero,
all $m_{\alpha}$ and $m_{\beta}$ dependence, and therefore the population
oscillations, completely disappear. 

We will therefore now concentrate on the distribution $F(\Lambda,\lambda)$
that appears in Eq.\ (\ref{eq4}): \begin{equation}
F(\Lambda,\lambda)=\left[\cos\Lambda+\cos\lambda\right]^{m_{1}}\left[\cos\Lambda-\cos\lambda\right]^{M-m_{1}}\end{equation}
 and study its variations as a function of the two variables, $\lambda$
and $\Lambda$. As in section \ref{Fockphase}, the band near $\Lambda=0$
will be called the {}``classical region'', the rest of the $\lambda$,$\Lambda$
plane the {}``quantum region''.

By taking the derivatives of the function $F(\Lambda,\lambda)$, we
find %
\footnote{$\Lambda=\pi$ does not occur because we have eliminated the redundant
regions beyond $\pi/2\le\Lambda\le3\pi/2$ from the integral of Eq.\ (\ref{eq4}).%
} that the peaks occur at \begin{eqnarray}
\Lambda_{m} & = & 0\quad\mathrm{and}\quad\lambda_{m}=\pm2\arctan\left(\sqrt{\frac{m_{2}}{m_{1}}}\right)\label{Peaks1}\end{eqnarray}
 \begin{eqnarray}
\Lambda_{m} & = & \pm2\arctan\left(\sqrt{\frac{m_{1}}{m_{2}}}\right)\quad\mathrm{and}\quad\lambda_{m}=0\label{Peaks2}\end{eqnarray}
 \begin{equation}
\Lambda_{m}=\pm2\arctan\left(\sqrt{\frac{m_{2}}{m_{1}}}\right)\quad\mathrm{and}\quad\lambda_{m}=\pi\label{Peaks3}\end{equation}

The peaks given by (\ref{Peaks1}) fall in the classical region, and
their position depends on the observed ratio between $m_{1}$ and
$m_{2}$; this is expected classically since the ratio of the two
intensities at the interferometer depends on the relative phase of
the two inputs. The other peaks fall in the quantum region, and will
be studied graphically in the next subsection.

\subsubsection{Graphical discussion; population oscillations}

\begin{figure}[t]
\includegraphics[width=3in]{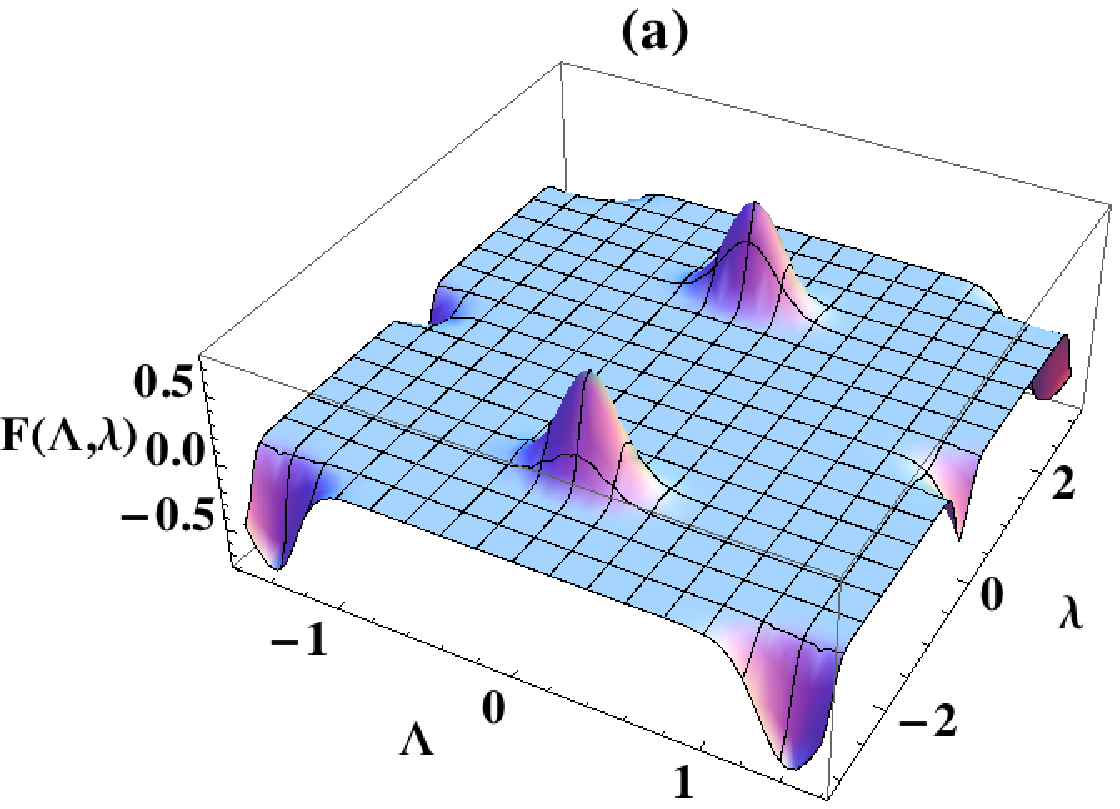} \hspace{0.2in} \includegraphics[width=3in]{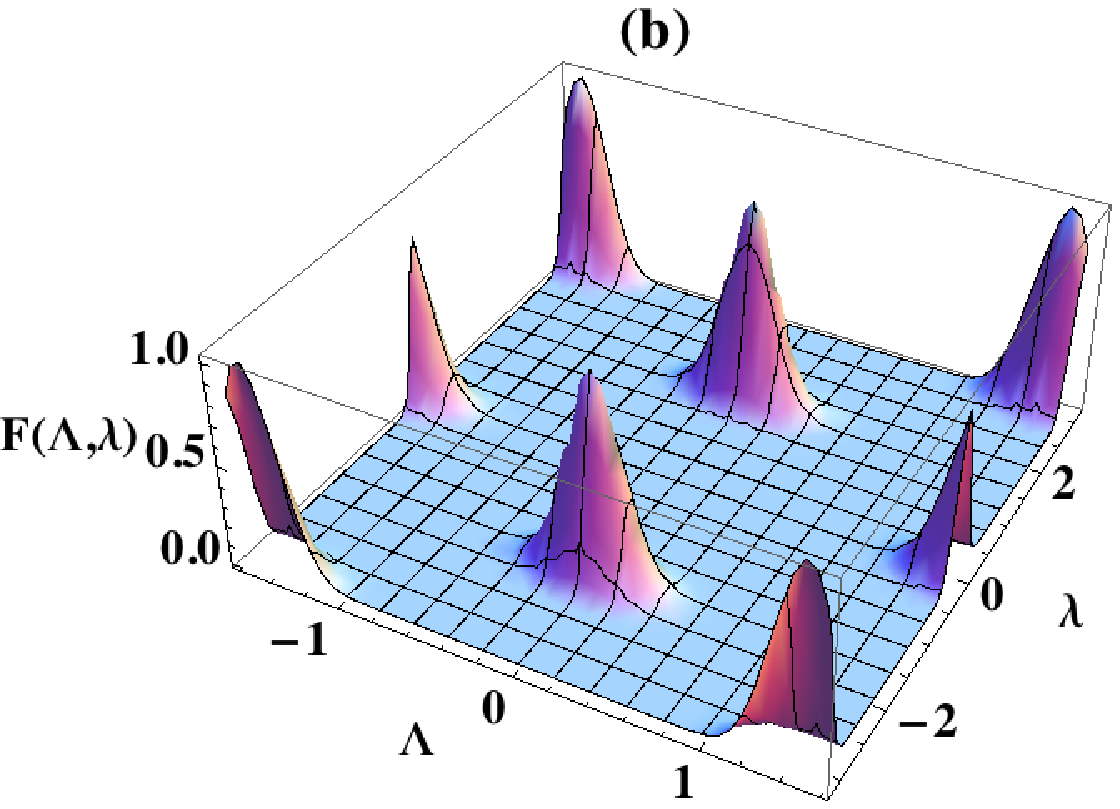}
\caption{(Color online) Plots of $F(\Lambda,\lambda)$ as a function of $\Lambda$ and $\lambda$
for $N_{\alpha}=N_{\beta}=M=40$ with: (a) $m_{1}=17$ and $m_{2}=23$;
(b) $m_{1}=18$ and $m_{2}=22$. Note that $-\pi\leq\lambda\leq\pi$
while $-\pi/2\leq\Lambda\le\pi/2$. In the {}``classical region''
($\Lambda\sim0$ ) the two phase peaks have a position that depends
on the ratio $m_{1}/m_{2}$, as expected classically. The peaks at
the corners of the quantum region change sign with the parity of $m_{1}$
and $m_{2}$; they are the source of the populations oscillations
shown in Fig. \ref{FigP1010}}

. \label{F(Lamlam)1010} 
\end{figure}

We make plots of the quantity $F(\Lambda,\lambda)$ by assuming that
$N_{\alpha}=N_{\beta}=M=40$. The multiple peaks are visible in Fig.
\ref{F(Lamlam)1010} for $m_{1}=17$ and $m_{2}=23$, as well as $m_{1}=18$
and $m_{2}=22$. The peaks in the figures occur, for $m_{1}=17$,
at $(\Lambda,\lambda)=(0,\pm1.72),(\pm1.72,0),(\pm1.42,\pm\pi)$ and,
for $m_{1}=18$, at $(\Lambda,\lambda)=(0,\pm1.67),(\pm1.67,0),(\pm1.47,\pm\pi)$.
The two first peaks in the {}``classical region'' correspond to
Eq. (\ref{Peaks1}), while all the other fall in the {}``quantum
region''. Because peaks corresponding to Eqs.\ (\ref{Peaks2}) and
(\ref{Peaks3}) add up at the border of the diagram, for these values
of the variables the four quantum peaks at the corners have positions
near $\pi/2$; they are therefore almost independent of the ratio
$m_{1}/m_{2}$ (if we had chosen smaller $m_{1}$ values, these peaks
would nevertheless have moved inside $\pi/2$), in contrast with the
classical peaks. Moreover, they have a sign that depends on the parity
of $m_{1}$ and $m_{2}$, so that it is clear that the two kinds of
peaks behave rather differently.

\begin{figure}[t]
 \includegraphics[width=3in]{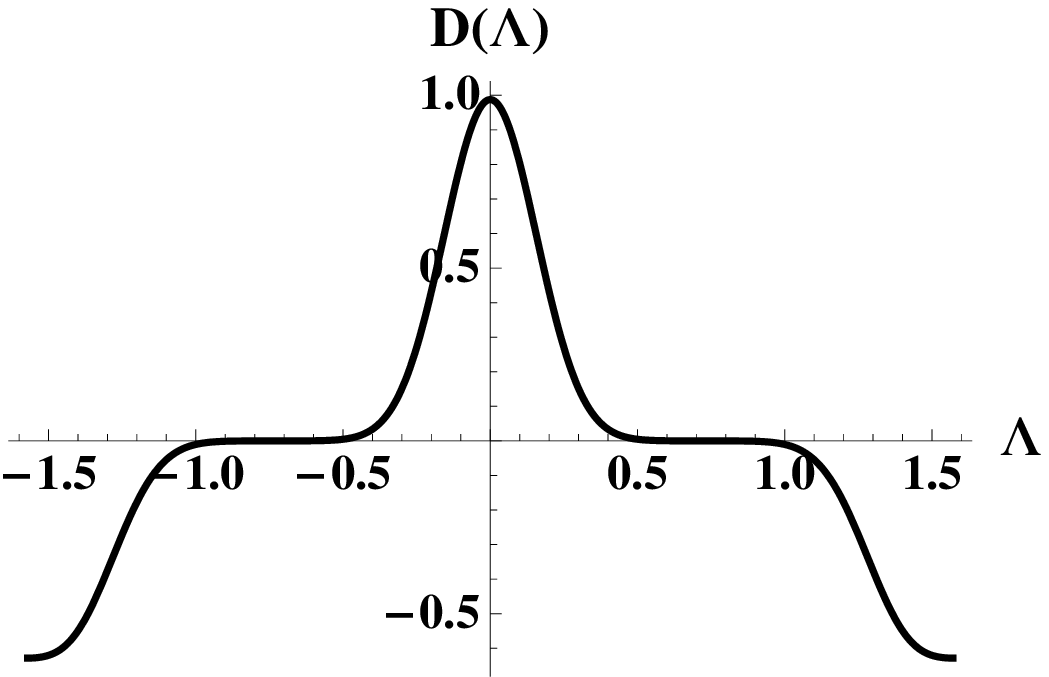} \hspace{0.2in} \includegraphics[width=3in]{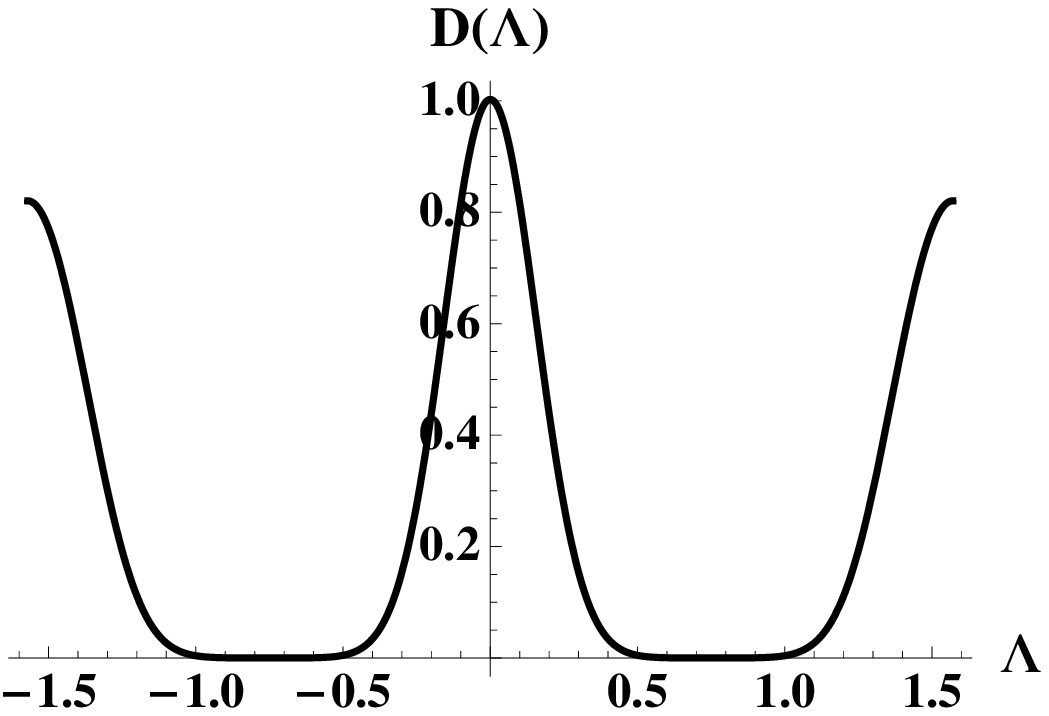}
\caption{Plots of $D(\Lambda),$ Eq.\ (\ref{D(Lam)}), the integral over $\lambda$
of the function shown in Fig. \ref{F(Lamlam)1010} for $N_{\alpha}=N_{\beta}=M=40$:
(a) m$_{1}=17,m_{2}=23$ and (b) $m_{1}=18,m_{2}=22.$}

\label{LamLine1010} 
\end{figure}

From Eq.\ (\ref{eq4}) we can obtain a quantum angle $\Lambda$ distribution
given by integrating $F$: \begin{equation}
D(\Lambda)=\int_{-\pi}^{\pi}\frac{d\lambda}{2\pi}\left[\cos\Lambda+\cos\lambda\right]^{m_{1}}\left[\cos\Lambda-\cos\lambda\right]^{M-m_{1}}\label{D(Lam)}\end{equation}
 Here the distribution $D(\Lambda)$ has two peaks, as shown in Fig.
\ref{LamLine1010}; these peaks are, via the Fourier cosine transform,
the source of the {}``population oscillations'' as a function of
$m_{\alpha}$.

Suppose for the moment that the peaks in $D(\Lambda)$ were $\delta$-functions
at $\Lambda=0$ and $\pi/2;$ then the cosine transform would be \begin{eqnarray}
P(m_{\alpha},m_{\beta}) & = & \int_{-\pi/2}^{\pi/2}d\Lambda\cos[(m_{\alpha}-m_{\beta})\Lambda]\left[\delta(\Lambda)\pm\delta(\Lambda+\pi/2)\pm\delta(\Lambda-\pi/2)\right]\nonumber \\
 & = & 1\pm\cos[(m_{\alpha}-m_{\beta})\pi/2]\label{cosinePO}\end{eqnarray}
 which oscillates with $m_{\alpha}$ as we have claimed in the form
of Eq. (\ref{eq:QualitativePOForm}). Whether the pattern has a maximum
or a zero at $m_{\alpha}=m_{\beta}$ depends on whether $m_{\alpha}$
is odd or even.

\begin{figure}[t]
 \includegraphics[width=3in]{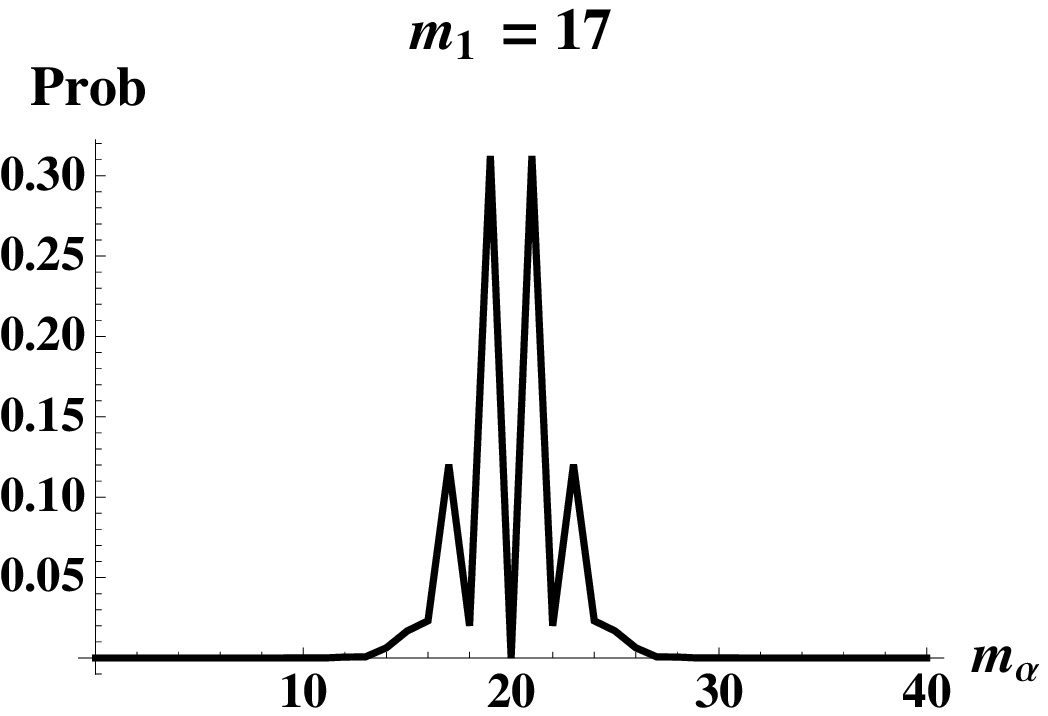} \hspace{0.2in} \includegraphics[width=3in]{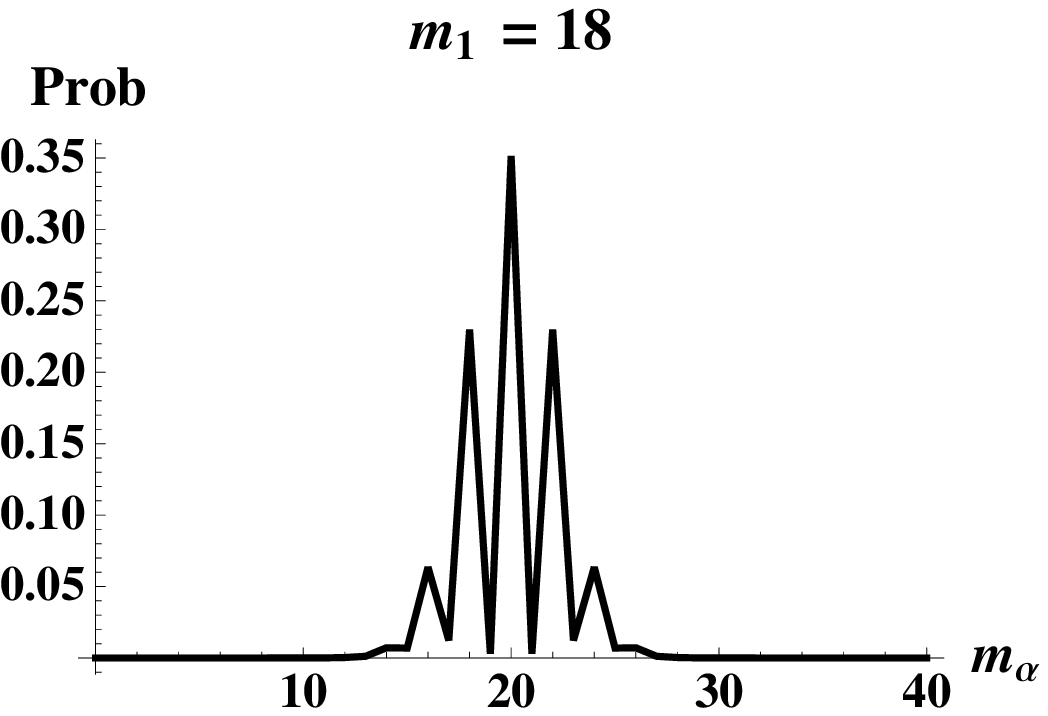}
\caption{Plots of $P(m_{1},m_{\alpha})$ of Eq.\ (\ref{Prob2Cond}) or Eq.\ (\ref{eq4})
versus $m_{\alpha}$, for $N_{\alpha}=N_{\beta}=40$ and (a) $m_{1}=17,m_{2}=23$
and (b) $m_{1}=18,m_{2}=22$. Only the integer values of $m_{\alpha}$
are relevant; the linear interpolation between them is just a guide
for the eye. Here $N_{\alpha}=N_{\beta}=M=40$}

\label{FigP1010} 
\end{figure}

The actual plots of $P(17,23,m_{\alpha},40-m_{\alpha})$ and
$P(18,22,m_{\alpha},40-m_{\alpha})$ are shown in Fig.  \ref{FigP1010};
the probability distribution in each case has a finite width, in
contrast to the distribution shown in Eq.\ (\ref{cosinePO}), because
of the finite width of the peaks in shown in $D(\lambda)$.  The shift
in phase of the two plots (one vanishing in the middle and the other
having a maximum) shows that the two components of the interference
have changed sign from one case to the other.  This is precisely the
case of the peaks in the quantum region in Fig.  \ref{F(Lamlam)1010}.
Moreover, the period of oscillation is constant (maximal for one value
of the population and minimal for the next), independent of the ratio
$m_{1}/m_{2}$, and therefore of the position of the peaks in the
classical regions.  These curves show the results corresponding to
the measurements of all four quantities $m_{1}, m_{2} ,\cdots$, in other words to
correlations between various measurements at the detectors.  If the
results are summed over $m_{1}$ at constant sum $m_{1}+m_{2}$, clearly, the
oscillations wash out.  In practice, this means that a post-selection
procedure is necessary in the experiments.

\begin{figure}[t]
 \includegraphics[width=3in]{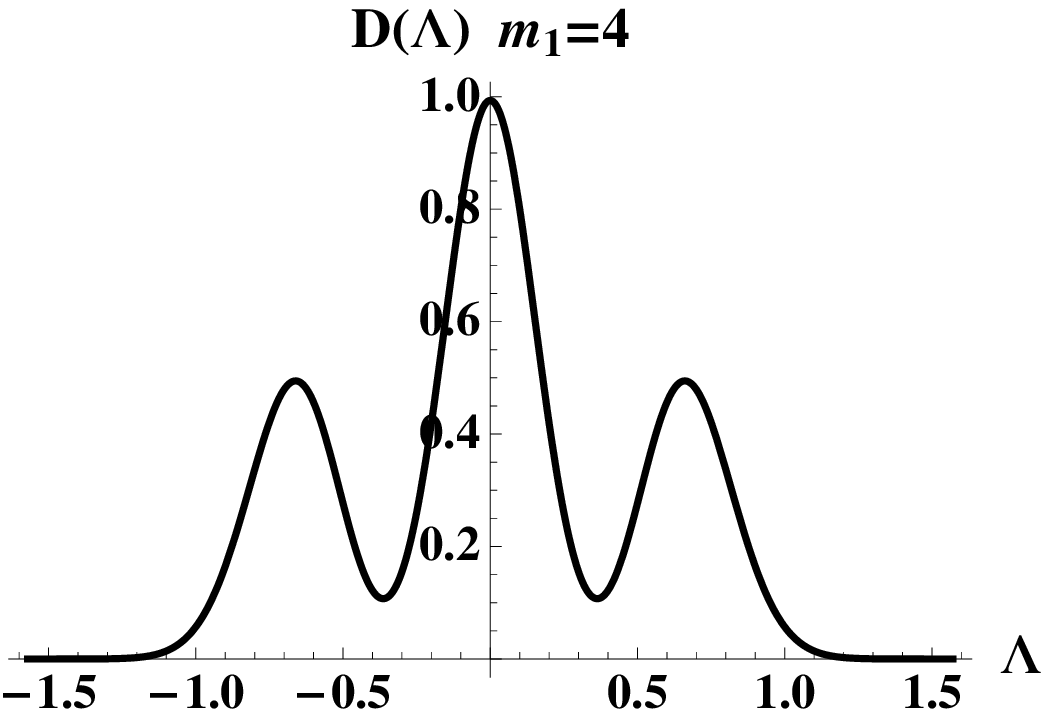} \hspace{0.2in} \includegraphics[width=3in]{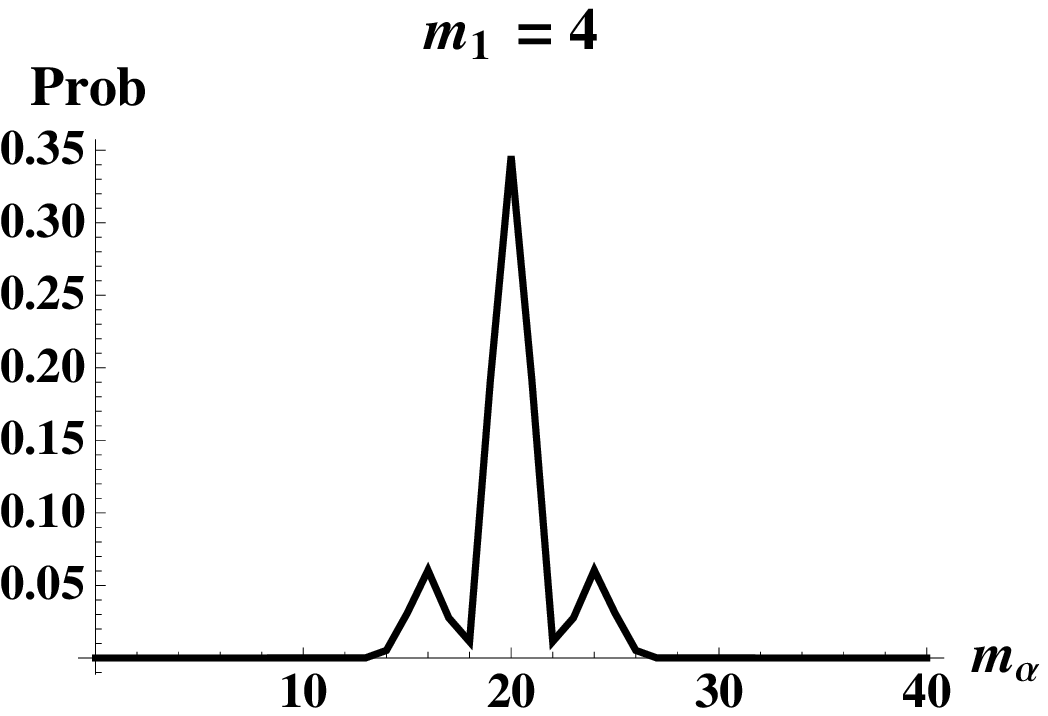}
\caption{Plot of (a) $D(\Lambda)$ and (b) $P(m_{1},m_{\alpha})$ for $N_{\alpha}=N_{\beta}=M=40$
and $m_{1}=4,m_{2}=36$. }

\label{m4DPO} 
\end{figure}

When $m_{1}\le16$ the outer peaks in $D(\Lambda)$ are no longer
positioned close to $\pi/2$ but move in to lower $\Lambda$ values,
and a minimum appears at $\pi/2$. Nevertheless, oscillations continue
to occur for values as small as $m_{1}=1$. Only at $m_{1}=0$ does
the population oscillation curve show just a single central peak.
As an example we show the case of $m_{1}=4$ in Fig. \ref{m4DPO}.

\begin{figure}[h]
\includegraphics[width=7cm]{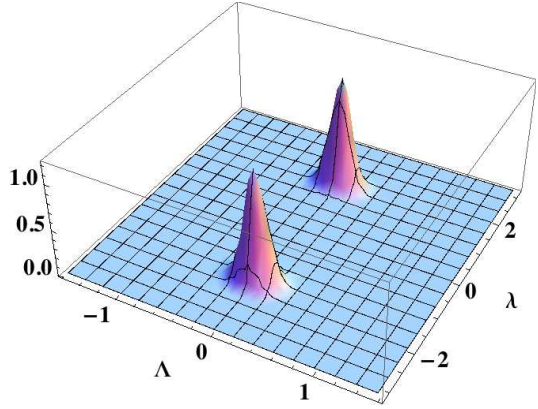}

\caption{(Color online) Plot of $(\cos\Lambda)^{N-M}F(\Lambda,\lambda)$ for $N_{\alpha}=N_{\beta}=40$
and $m_{1}=18,m_{2}=22$. Note that the peaks of Fig. \ref{F(Lamlam)1010}
that are away from $\Lambda=0$ are missing here.}

\label{ClassLamlam1010} 
\end{figure}

We finally discuss the $\lambda$ distribution. For this purpose,
we sum over variables $m_{\alpha},m_{\beta}$ to get a probability
of getting the distribution $\{m_{1},m_{2}\}$ independent of the
source distribution. To do the sum we must take into account the relation
$m_{\alpha}+m_{\beta}=N-M$ where $M=m_{1}+m_{2}$, with $M$ and
$N$ fixed. We obtain (see Appendix A) \begin{eqnarray}
P(m_{1},m_{2}) & = & \frac{N_{\alpha}!N_{\beta}!}{m_{1}!m_{2}!2^{N-1}}\int_{-\pi}^{\pi}\frac{d\lambda}{2\pi}\int_{-\pi/2}^{\pi/2}\frac{d\Lambda}{2\pi}e^{-i\left(N_{\alpha}-N_{\beta}\right)\Lambda}\sum_{m_{\alpha}}\frac{\left(e^{i\Lambda}\right)^{m_{\alpha}}\left(e^{-i\Lambda}\right)^{N-M-m_{\alpha}}}{m_{\alpha}!(N-M-m_{\alpha})!}\nonumber \\
 &  & \times\left[\cos\Lambda+\cos\lambda\right]^{m_{1}}\left[\cos\Lambda-\cos\lambda\right]^{m_{2}}\nonumber \\
 & = & \frac{N_{\alpha}!N_{\beta}!}{m_{1}!m_{2}!2^{M-1}}\int_{-\pi}^{\pi}\frac{d\lambda}{2\pi}\int_{-\pi/2}^{\pi/2}\frac{d\Lambda}{2\pi}\cos\left[\left(N_{\alpha}-N_{\beta}\right)\Lambda\right]\left(\cos\Lambda\right)^{N-M}\nonumber \\
 &  & \times\left[\cos\Lambda+\cos\lambda\right]^{m_{1}}\left[\cos\Lambda-\cos\lambda\right]^{m_{2}}\label{reducedInterf}\end{eqnarray}
 The distribution $F(\lambda,\Lambda)$ is now multiplied by $(\cos\Lambda)^{N-M}$,
which, for large $N-M$, peaks up sharply at $\Lambda=0$ and damps
out all peaks away from $\Lambda=0$, as shown in Fig. \ref{ClassLamlam1010};
the formula then reduces to the classical case.

\begin{figure}[h]
 \includegraphics[width=7cm]{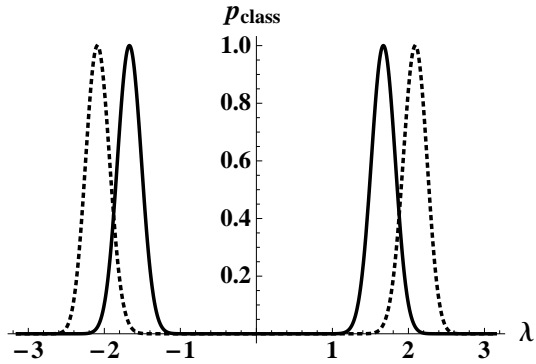}

\caption{Plot of $p_{class}(\lambda)$ of Eq.\ (\ref{Classp(lam)}), i.e.
the integral over $\Lambda$ of the function shown in Fig. \ref{ClassLamlam1010}
for $N_{\alpha}=N_{\beta}=M=40$. The solid line corresponds to $m_{1}=18$,
$m_{2}=22$ with peaks at $\pm1.67$; the dotted line is for $m_{1}=10$,
$m_{2}=30$ with peaks at $\pm2.09.$ As expected classically, the
peaks move symmetrically with change in $m_{1}$. }

\label{PlotClassplam} 
\end{figure}

The classical phase quasi-distribution is then \begin{equation}
p_{\mathrm{class}}(\lambda)=\int_{-\pi/2}^{\pi/2}\frac{d\Lambda}{2\pi}(\cos\Lambda)^{N-M}\left[\cos\Lambda+\cos\lambda\right]^{m_{1}}\left[\cos\Lambda-\cos\lambda\right]^{M-m_{1}}\label{Classp(lam)}\end{equation}
 A plot of this function for the same variable values is shown in
Fig. \ref{PlotClassplam}. Only the two classical peaks survive here.

An interesting feature of the PO is that, while within the reduced
probability of Eq.\ (\ref{reducedInterf}) one can replace $\Lambda$
by zero and get the classical limit, it is \emph{not correct} in Eq.
(\ref{eq4}), which contains no factor $(\cos\Lambda)^{N-M}$. The
result is that one can still get strong populations oscillations and
marked even-odd changes, even in the limit $M\ll N.$ The quantum
angle $\Lambda$ therefore remains necessary even in this case.

Population oscillations can continue to exist under certain circumstances
even if some particles are missed in the measurements; they are more
robust in this respect than violations of locality. This point is
discussed in Appendix A.

\subsection{Population oscillations with interference fringes in free space}

\begin{figure}[h]
 \centering \includegraphics[width=3in]{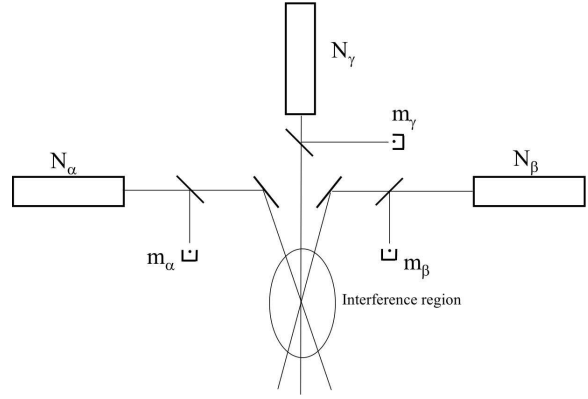} \caption{Particle beams from three sources emitting $N_{\alpha}$, $N_{\beta}$,
and $N_{\gamma}$ particles, respectively, interfere in free space
and can produce an interference pattern on a screen within the central
region. Some particles ($m_{\alpha}$, $m_{\beta}$, and $m_{\gamma}$)
are deflected near the sources to be counted in detectors. In our
simulation we have $N_{\alpha}=N_{\beta}=N_{\gamma}=N$; $N$ particles
reach the interference region and $m_{\alpha}+m_{\beta}+m{\gamma}=2N$
particles are deflected near the sources.}

\label{DBRPDevice} 
\end{figure}

We now attempt to reproduce the analysis of Dunningham et al. (DBRP)
in Ref.\ \cite{Dunn} in which three Fock sources form an interference
pattern in free space on a screen, while some of the particles are
deflected near the sources by beam splitters, where they are counted.
Fig. \ref{DBRPDevice} shows the experimental arrangement considered.
We will designate $M$ as the number of particles involved in the
interference measurements made on the screen where interference takes
place. Then the number of particles measured near the sources having
initial particle numbers $N_{\alpha}=N_{\beta}=N_{\gamma}=N$ (as
in the work of DBRP) will be $m_{\alpha}$, $m_{\beta},$ and $m_{\gamma}$;
these are the particles that did not take part in the interference
pattern. All together then we will have measured \begin{equation}
3N=M+m_{\alpha}+m_{\beta}+m_{\gamma}\label{3N}\end{equation}
 particles. We can then write the probability as\begin{equation}
P(m_{\alpha},m_{\beta},m_{\gamma},r_{1},\cdots,r_{M})\sim\left\langle \Gamma_{m_{\alpha}m_{\beta}m_{\gamma}M}\left|\Gamma_{m_{\alpha}m_{\beta}m_{\gamma}M}\right.\right\rangle \end{equation}
 where \begin{equation}
\left|\Gamma_{m_{\alpha}m_{\beta}m_{^{\gamma}}M}\right\rangle =\frac{a_{\alpha}^{m_{\alpha}}a_{\beta}^{m_{\beta}}a_{\gamma}^{m_{\gamma}}}{\sqrt{m_{\alpha}!m_{\beta}!m_{\gamma}!}}\prod_{i=1}^{M}(a_{\alpha}e^{ik_{\alpha}\cdot r_{i}}+e^{ik_{\beta}\cdot r_{i}}a_{\beta}+a_{\alpha}e^{ik_{\gamma}\cdot r_{i}})\left|N,N,N\right\rangle \label{Gamma2}\end{equation}
 To correspond with Ref.\ \cite{Dunn} we take $k_{\alpha}=k,$ $k_{\beta}=-k$
and with $k_{\gamma}=0$. %

We can introduce a vacuum state in between the $\left|\Gamma\right\rangle $'s
and compute the matrix element by multiplying out the interference
operators:\begin{eqnarray}
\left\langle 0\left|\right.\Gamma\right\rangle  & = & \frac{1}{\sqrt{m_{\alpha}!m_{\beta}!m_{\gamma}!}\sqrt{N!^{3}}}\sum_{p}K_{p_{a}p_{\beta}p_{\gamma}}(r)\left\langle 0\left|a_{\alpha}^{m_{\alpha}+p_{\alpha}}a_{\beta}^{m_{\beta}+p_{\beta}}a_{\gamma}^{m_{\gamma}+p_{\gamma}}a_{\alpha}^{\dagger N}a_{\beta}^{\dagger N}a_{\gamma}^{\dagger N}\right|0\right\rangle \end{eqnarray}
 where $K_{p}$ is a coefficient that depends on the $r_{i}$. The
matrix element produces delta functions that can be replaced by integrals
in our standard way. The results is \begin{eqnarray}
\left\langle 0\left|\right.\Gamma\right\rangle  & = & \frac{\sqrt{N!^{3}}}{\sqrt{m_{\alpha}!m_{\beta}!m_{\gamma}!}}\int\frac{d\lambda_{\alpha}}{2\pi}\int\frac{d\lambda_{\beta}}{2\pi}e^{-i(N_{\alpha}-m_{\alpha})\lambda_{\alpha}}e^{-i(N_{\beta}-m_{\beta})\lambda_{\alpha}}\\
 &  & \prod_{i=1}^{M}(e^{ik\cdot r_{i}}e^{i\lambda_{\alpha}}+e^{-ik\cdot r_{i}}e^{i\lambda_{\beta}}+1)\nonumber \end{eqnarray}

If we take the absolute square of this we introduce two new variables
$\lambda_{\alpha}^{\prime}$ and $\lambda_{\beta}^{\prime}$ . We
then make the following variable changes:\begin{eqnarray}
\lambda_{\alpha} & = & -\lambda+\frac{\Lambda}{2},\quad\lambda_{\alpha}^{\prime}=-\lambda-\frac{\Lambda}{2},\nonumber \\
\lambda_{\alpha} & = & \lambda^{\prime}+\frac{\Lambda^{\prime}}{2},\quad\lambda_{\beta}=\lambda+\frac{\Lambda^{\prime}}{2},\label{Vars}\end{eqnarray}
 The probability then becomes \begin{eqnarray}
P(m_{\alpha},m_{\beta},m_{\gamma},r_{1},\cdots,r_{M}) & = & \frac{N!^{3}}{m_{\alpha}!m_{\beta}!m_{\gamma}!}\int_{-\pi}^{\pi}\frac{d\lambda}{2\pi}\int_{-\pi}^{\pi}\frac{d\lambda^{\prime}}{2\pi}\int_{-\pi}^{\pi}\frac{d\Lambda}{2\pi}\int_{-\pi}^{\pi}\frac{d\Lambda^{\prime}}{2\pi}e^{-i(N-m_{\alpha})\Lambda}\nonumber \\
 &  & \times e^{-i(N-m_{\beta})\Lambda^{\prime}}\prod_{i=1}^{M}\left[1+e^{i\Lambda}+e^{i\Lambda^{\prime}}\right.\nonumber \\
 &  & +2\cos(2k\cdot r_{i}-\lambda-\lambda^{\prime})e^{i(\Lambda+\Lambda^{\prime})/2}\nonumber \\
 &  & +\left.2\cos(k\cdot r_{i}-\lambda)e^{i\Lambda/2}+2\cos(k\cdot r_{i}-\lambda^{\prime})e^{i\Lambda^{\prime}/2}\right]\label{3CProbIntegral}\end{eqnarray}
 If we sum out the $m_{i}$ we find the probability for the $r-$set
under arbitrary source number detections:

\begin{eqnarray}
P(r_{1},\cdots,r_{M}) & = & \frac{N!^{3}}{(3N-M)!}\int_{-\pi}^{\pi}\frac{d\lambda}{2\pi}\int_{-\pi}^{\pi}\frac{d\lambda^{\prime}}{2\pi}\int_{-\pi}^{\pi}\frac{d\Lambda}{2\pi}\int_{-\pi}^{\pi}\frac{d\Lambda^{\prime}}{2\pi}e^{-iN(\Lambda+\Lambda)}\nonumber \\
 &  & \times\left(1+e^{i\Lambda}+e^{i\Lambda^{\prime}}\right)^{3N-M}\nonumber \\
 &  & \times\prod_{i=1}^{M}\left[1+e^{i\Lambda}+e^{i\Lambda^{\prime}}+2\cos(2k\cdot r_{i}-\lambda-\lambda^{\prime})e^{i(\Lambda+\Lambda^{\prime})/2}\right.\nonumber \\
 &  & +\left.2\cos(k\cdot r_{i}-\lambda)e^{i\Lambda/2}+2\cos(k\cdot r_{i}-\lambda^{\prime})e^{i\Lambda^{\prime}/2}\right]\label{3CReducedProbInt}\end{eqnarray}

\begin{figure}[h]
 \centering \includegraphics[width=3in]{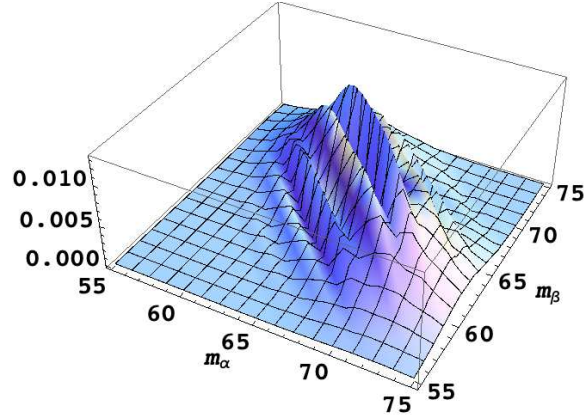}

\caption{(Color online) The probability of measurement of populations, made with beam splitters
near the three sources for $M=100$ position measurement in the interference
pattern in free space for three-condensates, where initially each
source had $N=100$ particles The two horizontal axes are the source
populations $m_{\alpha}$ and $m_{\beta}$ of the condensates with
opposite wave vectors; the source count $m_{\gamma}$, corresponding
to the condensate with zero wave vector, is equal to $50-m_{\alpha}-m_{\beta}$. }

\label{Ridges} 
\end{figure}

The integral method above is not very useful for simulations. We have
developed a recurrence method in which the wave function for $R$
measurements is written in terms of that for $R-1$ measurements.
We do not give details here to save space. All the probabilities $p(m_{\alpha},m_{\beta})$
for finding particles in the source detectors for a given set of positions
$r_{1},\cdots,r_{M}$ (with $m_{\gamma}$ given by Eq.\ (\ref{3N}))
are computed in a single recurrence run. We show a plot in Fig. \ref{Ridges}
of the resulting population oscillation ridges. DBRP found that the
ridges were parallel to one axis shown in their Fig. 5. Indeed our
ridges are parallel to $m_{\gamma}=$constant, which would have been
more obvious had we plotted using, say, the $m_{\alpha},m_{\gamma}$
axes. The parallel axis in our case is the one having the intermediate
vector $(k_{\gamma}=0)$ so that we are in agreement with the results
of DBRP. %

\section{Conclusion}

In many cases, such as for instance the description of the MIT experiment
\cite{WK} with initial Fock states, introducing a classical relative
phase angle $\lambda$ is sufficient. Cases exist, nevertheless, where
the classical phase is not able to explain all quantum predictions,
and where introduction of the quantum angle $\Lambda$ (or its equivalent)
becomes necessary. We have discussed two examples, in Secs. \ref{Interferometeranalysis}
and \ref{Populationoscillations}, where interesting physical effects
can not be understood only in terms of the classical phase. In both
cases, quantum effects are related to peaks of the function $F(\lambda,\Lambda)$
in the {}``quantum region'' (i. e., away from $\Lambda=0$), and
disappear completely if $\Lambda$ is set to zero.

In the first double interferometer experiment, we find violations
of the BCHSH inequalities and therefore violations of locality. Setting
the quantum angle to zero reduces the equations to purely classical
equations, which could be interpreted as being integrated over a hidden
variable as in Bells theorem. Only the quantum angle leads to the
violations.

In the population oscillation experiment, we find that simultaneous
measurements of {}``non-commuting variables'' phase and particle
number within the same apparatus yield oscillations in measurements
of the number variable that are a direct result of the off-diagonal
phase (i. e., quantum) peaks and provide an example of QIMDS. However,
as discussed in Appendix B, one can replace the measurement of the
phase by that of the parity, which does not fix the relative phase
of the two condensates at all; but this does not completely cancel
the population oscillations since the central dark fringe remains
present with a 100\% contrast, while the characteristics of the superposition
are completely changed (the {}``phase cat'' becomes completely {}``blurred'').
The fact that some population oscillations remain visible, at least
for the first fringe, illustrates that the PO can exist more generally
than with just the coherent superpositions of different macroscopic
phases.

The two experiments we have discussed are of somewhat different nature.
The former exhibits strong quantum non-locality effects, while for
the latter we have not found violations of the Bell inequalities.
Nevertheless, while for the former the violations of the inequalities
require that all particles are measured (they disappear as soon as
a single particle is missed), the population oscillations are a manifestation
of the quantum angle that is more robust as we show in Appendix A;
they can still exist, although in a more limited way, when a few particles
are missed. 
\begin{acknowledgments}
We wish to thank the authors of Ref.\ \cite{Dunn}, J.\ A. Dunningham,
K. Burnett, R.\ Roth and W.\ D.\ Phillips, as well as A. Smerzi
and A. Piazza, for interesting and helpful comments and discussions. 
\end{acknowledgments}
\vspace{20mm}

\begin{center}
APPENDICES 
\par\end{center}

\subsection{Incomplete measurements in the PO experiment}

\subsubsection{No phase measurements}

We study the experiment of Fig.\ \ref{PO interferometer} again,
but now assume that no measurement is performed in the interference
region D, and that only the population measurements are performed;
then, whether or not a beam splitter is used in this regions does
not matter anymore). We then have to sum the probabilities (\ref{eq4})
over $m_{1}$ and $m_{2}$, with a constant sum $m_{1}+m_{2}=M$.
The summation introduces the $M$-th power of a binomial $[e^{i\Lambda}+e^{-i\Lambda}]$,
but only one term of this power survives the $\Lambda$ integration;
we then obtain \begin{equation}
\sum_{m_{1}+m_{2}=M}P(m_{1},m_{2},m_{\alpha},m_{\beta})=\frac{N_{\alpha}!N_{\beta}!}{m_{\alpha}!m_{\beta}!\,2^{N-M}}\frac{M!}{p!(N-p)!}\label{eq4-bis}\end{equation}
 with $p$ defined by \begin{equation}
2p=N_{\alpha}-m_{\alpha}-N_{\beta}+m_{\beta}\label{eq4-ter}\end{equation}
 (one can easily check that the right hand side of this equation is
an even number). The $\lambda$ integral has now disappeared, as expected
since no measurement of the relative phase is made. Moreoever, the
probability factorizes as expected since, in the absence of interference
measurements, two completely independent experiments are performed
in different regions of space: in each region, the transmission or
reflection of the particles on the beam splitter are independent random
processes.

\subsubsection{No population measurements}

Conversely, assume that all population measurements are ignored and
that only the interference measurements are considered. The corresponding
probability is then \begin{eqnarray}
\sum_{m_{\alpha}+m_{\beta}=N-M}P(m_{1},m_{2},m_{\alpha},m_{\beta}) & = & \frac{N_{\alpha}!N_{\beta}!}{(N-M)!m_{1}!m_{2}!2^{M}}\int_{-\pi}^{\pi}\frac{d\lambda}{2\pi}\int_{-\pi}^{\pi}\frac{d\Lambda}{2\pi}\,[\cos\Lambda]^{N-M}\nonumber \\
 &  & \times\left[\cos\Lambda+\cos\lambda\right]^{m_{1}}\left[\cos\Lambda-\cos\lambda\right]^{m_{2}}\label{eq4-quater}\end{eqnarray}
 Now the phase $\lambda$ no longer disappears, but combines its effects
with the quantum angle $\Lambda$; the $[\cos\Lambda]^{N-M}$ introduces
a peaking function around the origin, which may behave similarly to
a delta function if $N-M$ is sufficiently large. We now discuss the
interplay between the classical phase and the quantum angle $\Lambda$.

\subsubsection{Missed particles}

\begin{figure}[h]
\centering \includegraphics[width=3in]{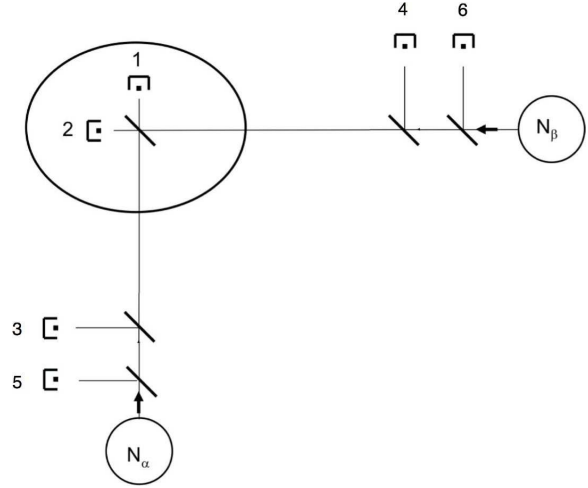}

\caption{Modified population device to show the effect of particles missed
in the measurement. The missed particles are supposed to enter detectors
5 and 6. We assume we know that $m_{5}+m$$_{6}=M_{L}$ particles
are missed, but we do not know their distribution in the new side
detectors. Thus we sum over all $m_{5},m_{6}$ to get the resulting
probability when some particles are not detected. }

\label{MoreSides} 
\end{figure}

Next suppose some of the particles are lost and not measured in either
interferomenter nor side detectors of Fig.\ \ref{PO interferometer}.
We have seen in the case of the double interferometer Bell-violation
experiment that a single missed particle can remove any locality violations.
We simulate these lost particles in the PO experiment by putting additional
side detectors as shown in Fig.\ \ref{MoreSides}. Assume that the
beam splitters at detectors 5 and 6 each have a transmission coefficient
$T.$ 

We assume that particle losses $m_{5}$ and $m_{6}$ are known to
total $M_{L}$, but the individual numbers are not actually recorded.
Thus to get the probability we are interested in we must sum over
all $m_{5}$ and $m_{6}$ adding to the total $M_{L}$. Proceeding
as in Sec. IV we find\begin{eqnarray}
P(m_{1},m_{2},m_{\alpha},m_{\beta}) & = & \frac{N_{\alpha}!N_{\beta}!T^{N-M_{L}}(R)^{M_{L}}}{m_{1}!m_{2}!m_{\alpha}!m_{\beta}!2^{N-2M_{L}}M_{L}!}\int_{-\pi}^{\pi}\frac{d\lambda}{2\pi}\int_{-\pi}^{\pi}\frac{d\Lambda}{2\pi}\cos\left[\left(N_{\alpha}-m_{\alpha}-N_{\beta}+m_{\beta}\right)\Lambda\right]\nonumber \\
 &  & \times\cos(\Lambda)^{M_{L}}\left[\cos\Lambda+\cos\lambda\right]^{m_{1}}\left[\cos\Lambda-\cos\lambda\right]^{m_{2}}\end{eqnarray}
with $R=1-T.$ The result of the lost particles is the factor $\cos(\Lambda)^{M_{L}},$
which, if $M_{L}$ is large enough, diminishes the quantum peaks,
as we have seen before as in, say, Eq. (\ref{eq4-quater}). The result
maintains the same form if we also allow particles to be missed elsewhere
in the device, say, after the beam splitter at detectors 1 and 2. 

If we count $N_{D}=m_{1}+m_{2}+m_{\alpha}+m_{\beta}$ particles in
the real detectors but $M_{L}$ were missed in one place or another
in the device, then we must have had $N=N_{D}+M_{L}$ particles in
the sources originally. The missed particles could have come from
source $\alpha$ or source $\beta$. We assume that the sources originally
have $N_{\alpha}=\bar{N_{\alpha}}+\Delta_{\alpha}$ and $N_{\beta}=\bar{N_{\beta}}+\Delta_{\beta}$
where $\bar{N_{\alpha}}+\bar{N_{\beta}}=N_{D}$ and $M_{L}=\Delta_{\alpha}+\Delta_{\beta}$.
We first fix $M_{L}$ and sum over all possible $\Delta_{\alpha}$
and then sum over all $M_{L}$ in principle from 0 to $\infty$. If
$R$ is small, then the sum over $M_{L}$ should converge after a
reasonably small number of missed particles. That is, the probability
of missing $X_{L}$ particles, where $X_{L}$ is very large is negligible.
One would hope that if $R$ is small enough, then the PO will converge
to a situation in which the fringes are not lost. We find this to
be the case under certain conditions. We can also find the average
number of particles lost by multiplying the probability by $M_{L}$
and summing over all $\Delta_{\alpha}$, $M_{L},$ and $m_{\alpha.}$

\begin{figure}[h]
$\quad$\includegraphics[width=3in]{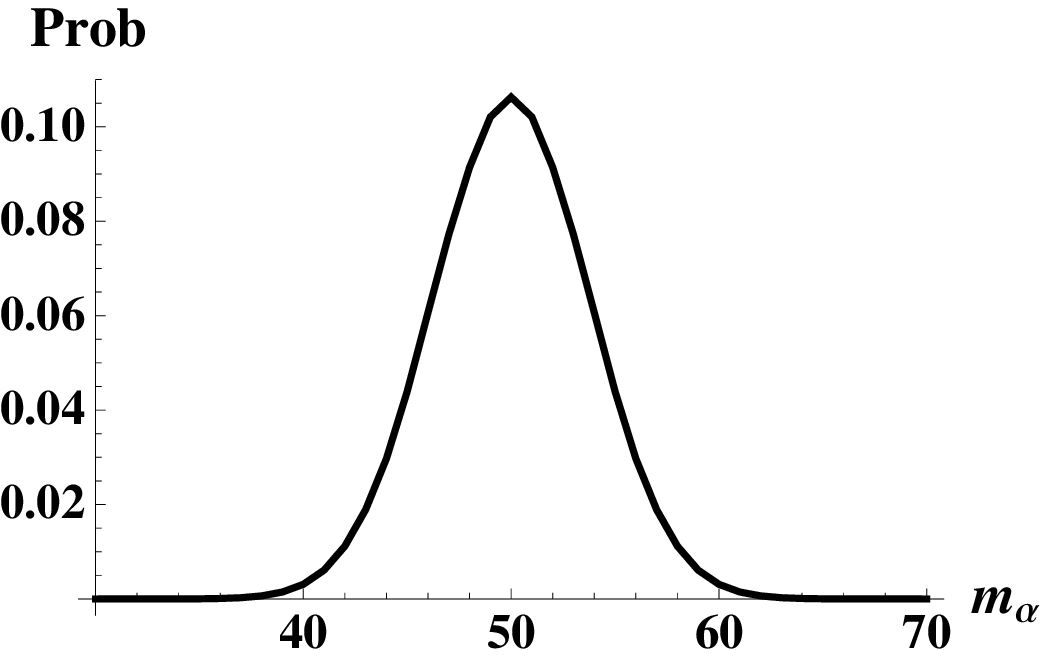}$\quad$\includegraphics[width=3in]{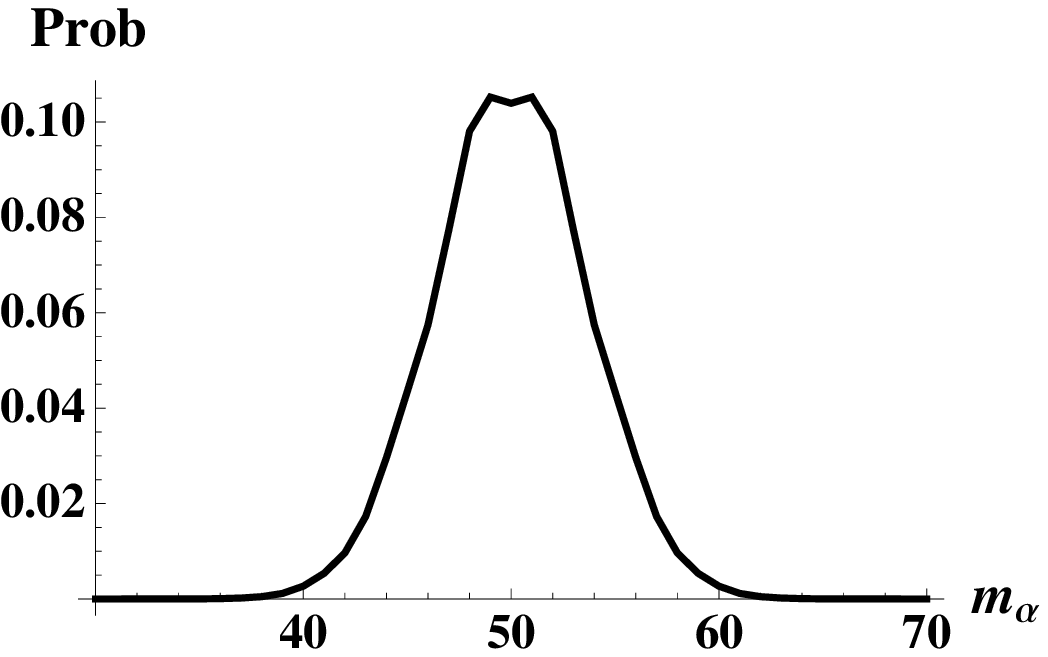}

\caption{The PO when the total number of detected particles in 200 with $m_{1}=17,$
$m_{2}=83$, and $T=0.98$ (left) and 0.99 (right). Note the very
small depression in the center of the right plot showing a remnant
of the PO after all possible losses are considered. The average numbes
of particles lost here are 6.5 and 3.8, respectively.}

\label{Missing 1}
\end{figure}

Consider the situation with $m_{1}=17$ and $m_{2}=83.$ The PO for
the cases with $T=0.98$ and $T=0.99$ are shown in Fig. \ref{Missing 1}.
For $T=0.98$ the oscillations are completely removed and for 0.99
only a remnant is left. In the later case we have lost 3.8 particles
on average. 

\begin{figure}[h]
$\quad$ \includegraphics[width=3in]{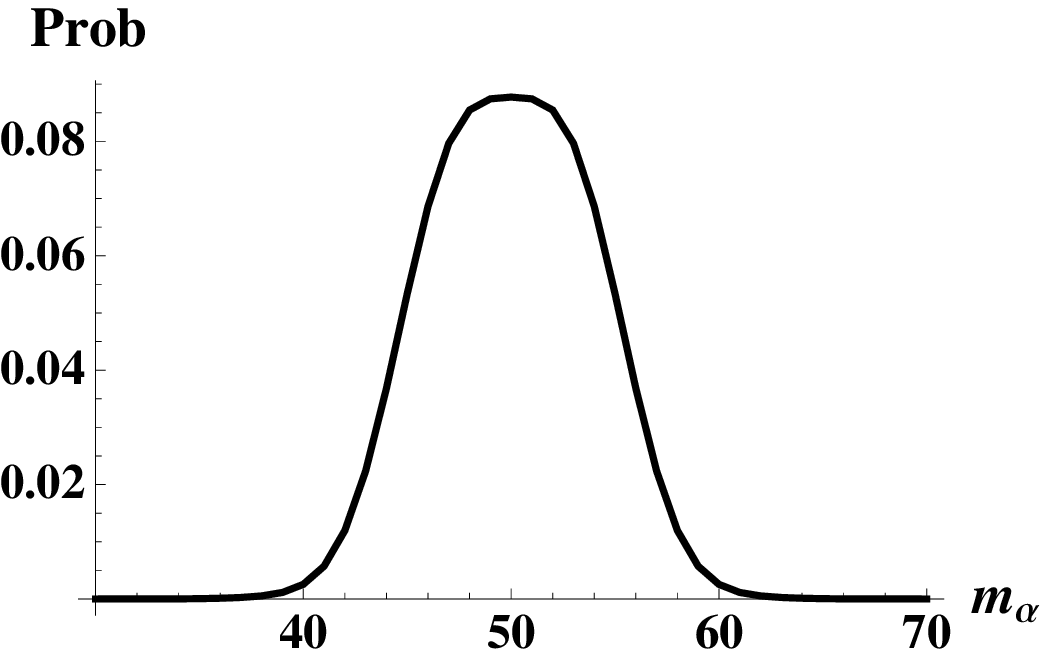}$\quad$\includegraphics[width=3in]{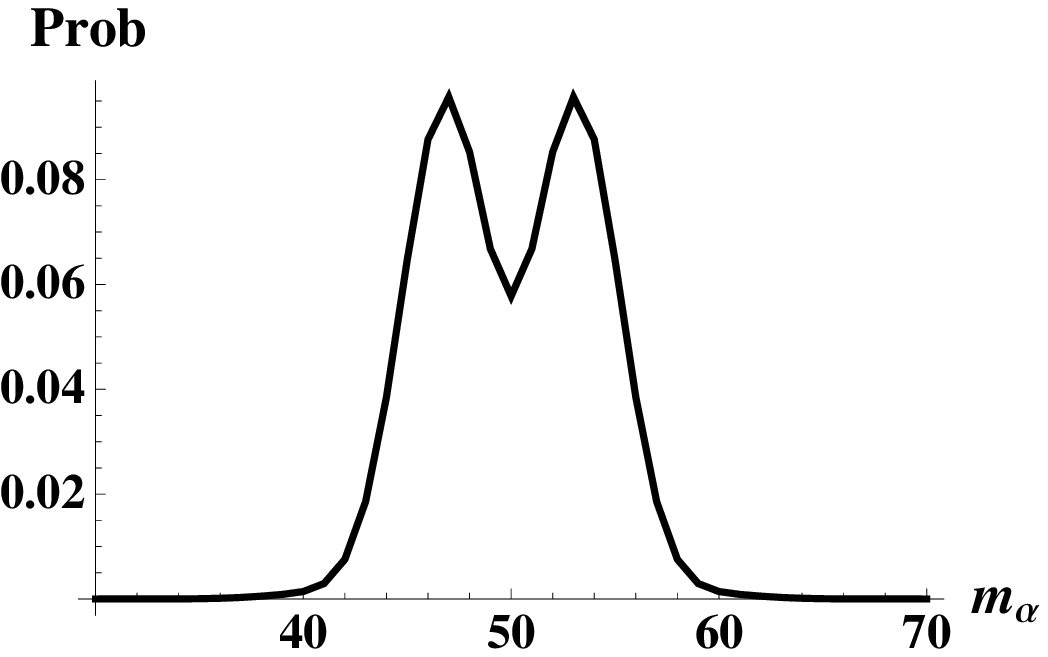}

\caption{The PO when the total number of detected particles in 200 with $m_{1}=3,$
$m_{2}=97$, and $T=0.98$ (left) and 0.99 (right). A much larger
remnant depression remains in the 0.99 case here. The average numbers
of particles lost here are again 6.5 and 3.8, respectively.}

\label{Missing 2}
\end{figure}

The smaller the value of $m_{1}$, the closer in to $\Lambda=0$ are
the off-diagonal peaks in $F(\Lambda,\lambda)$ so that they get less
blotted out by the $\cos\Lambda^{M_{L}}$ factor. If we lower $m_{1}$
to 3 we find the results in Fig. \ref{Missing 2}. We still lose the
central dip in the PO diagram for $T=0.98,$ but for $T=0.99,$we
get a much deeper remnant.


\begin{figure}[h]
\centering \includegraphics[width=3in]{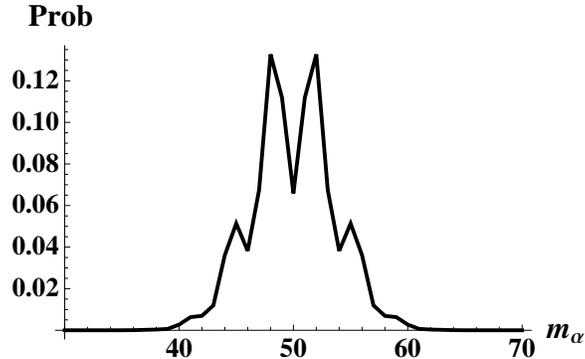}

\caption{The PO when the total number of detected particles in 200 with $m_{1}=17,$
$m_{2}=195$, and $T=0.997$ . Here we get more than just the central
dip. The average numbers of particles lost here is just 1.4.}

\label{Missing 3}
\end{figure}

To what degree must we restrict losses to guarantee that we would
have more than a single dip? Fig. \ref{Missing 3} shows the case
of $N=200,$ $m_{1}=17$ with the transmission coefficient up to $T=0.997.$
Only 1.4 particles have been lost here. For smaller $m_{1}$ values
one gets deeper central dips, but not the dips on the side for the
same $T$ values. %

\subsection{Measuring the parity}

Fig.\ \ref{F(Lamlam)1010} shows that the $\Lambda$ peaks in the
quantum region have a sign that depends on the parity of $m_{1}$
or $m_{2}$, which suggests that a possible method to observe the
population oscillations is to associate them with a measurement of
the parity at the interferometer, instead of the relative phase of
the two condensates.

\begin{figure}[t]
 \includegraphics[width=3in]{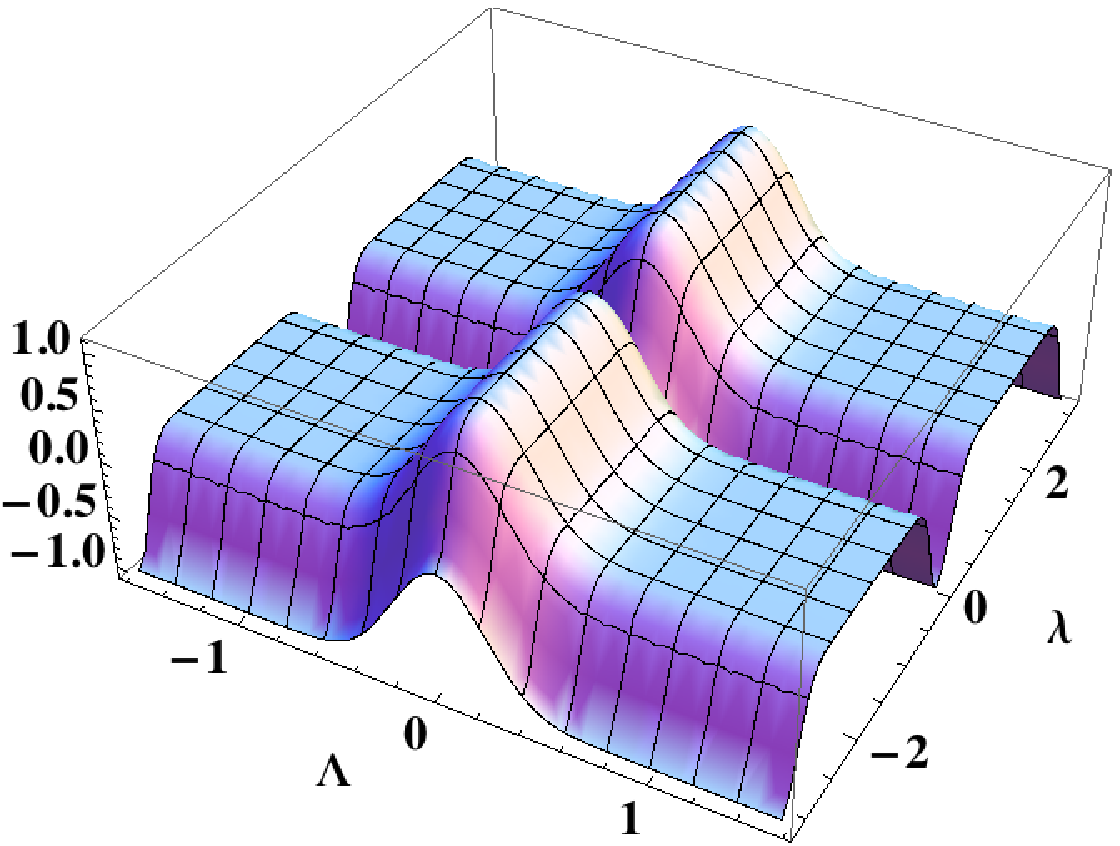} \hspace{0.2in} \includegraphics[width=3in]{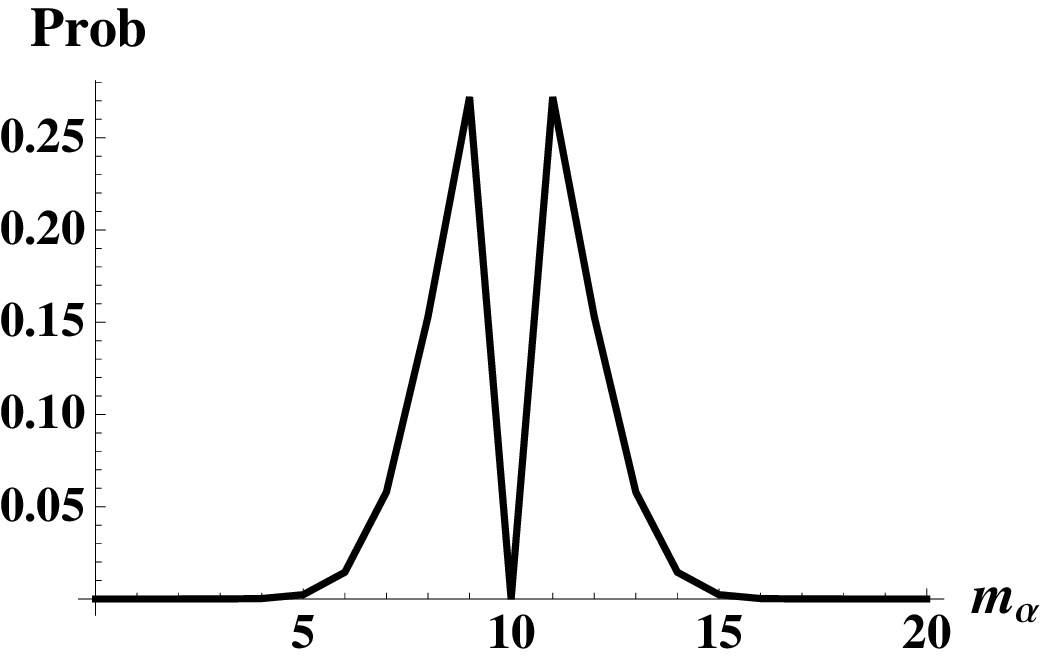}
\caption{(Color online) Left: plot of $F(\Lambda,\lambda)$ as a function of $\Lambda$ and
$\lambda$ for $N_{\alpha}=N_{\beta}=M=20$, obtained by summing the
probabilities of all odd values of $m_{1}$ and $m_{2}$. One notices
that the $\lambda$ peaks in the classical region are now spread over
many values of $\lambda$, except sharp variations around $\lambda=0$
and $\lambda=\pm\pi$; the function still takes significant values
in the quantum region $\Lambda\ne0$. Right: corresponding population
oscillations; the narrow central fringe is still perfectly visible
with a 100\% variation when $m_{1}$ varies by only one unit only
(the central fringe is dark because $m_{1}$ is odd).}

\label{Oddcats} 
\end{figure}

Fig.\ \ref{Oddcats} illustrates what is obtained if, for instance,
one adds the probabilities associated with all odd values of $m_{1}$.
The left part of the figure shows the variations of $F(\Lambda,\lambda)$,
the right part the associated population oscillation as a function
of $m_{1}$, with $m_{2}=40-m_{1}$. One notices the disappearance
of the two peaks that characterized the coherent superposition of
two values of the relative phase; they are now replaced by a more
delocalized structure, similar to a ridge. In other words, the {}``Schrödinger
cat'' is now spread over many values of the phase. But one also sees
in the right part that the populations oscillations still exist, with
a central dark fringe that has 100\% contrast when $m_{\alpha}$ varies
only by one unit; the variation is actually not very different from
the right part of Fig.\ \ref{m4DPO}, except of course the change
of sign due to the change of parity of $m_{1}$. This shows that the
central fringe of the population oscillations is not specifically
related to a measurement of the phase, or to the existence of any
{}``Schrödinger cat''; it continues to exist if a very different
physical quantity is measured, such as the parity, which does not
give any particular information on the relative phase of the two condensates.

\subsection{Two phase measurements}

\begin{figure}[h]
 \centering \includegraphics[width=3in]{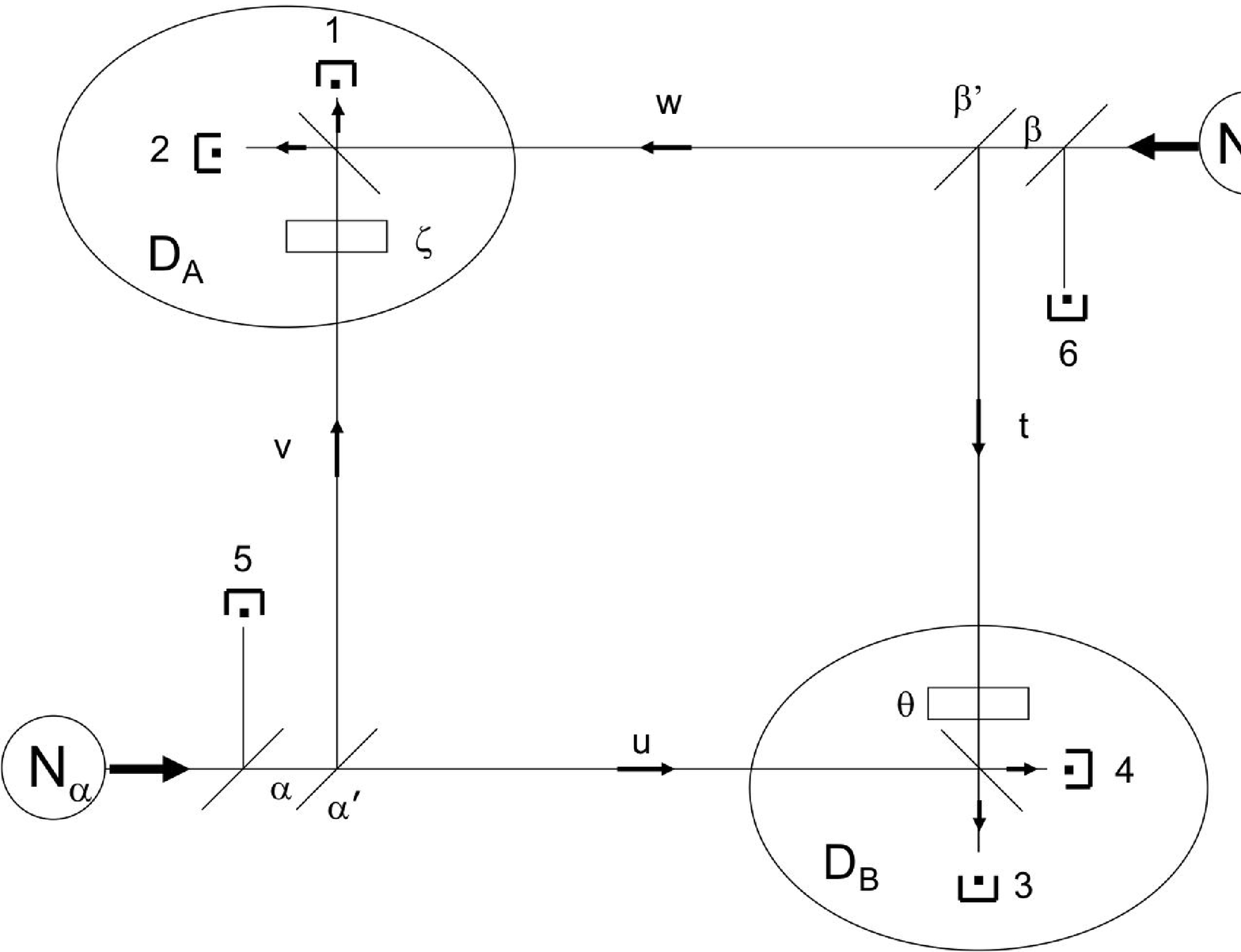}

\caption{Two Fock states, with populations $N_{\alpha}$ and $N_{\beta}$,
enter beam splitters, and are then made to interfere in two different
regions of space $D_{A}$ and $D_{B}$, with detectors 1 and 2 in
the former, 3 and 4 in the latter. In each of the channels $j=1,2,3,4$
particles are counted. The extra detector 5 and 6 count particles
that are not measured by the interferometers. }

\label{TwoInterfPO} 
\end{figure}

In the population oscillation experiment, we have considered the use
of a single interferometer; we can generalize this experiment to two
interferometers each of which have different settings, $\zeta$ and $\theta$.
We begin with the device in Fig.\ \ref{Interferometer} and add to
that the side detectors shown in Fig. \ref{PO interferometer}, to
allow phase-type measurements in two different regions of space as
well as population measurements near the two sources. The resulting
apparatus is shown in Fig. \ref{TwoInterfPO}.

The calculations proceed as in previous sections and lead to a result
for the probability of finding the series of results $\{m_{1},m_{2},m_{3},m_{4},m_{\alpha},m_{\beta}\}$
equal to 
\begin{eqnarray}
\mathcal{P}(m_{1},m_{2},m_{3},m_{4},m_{\alpha},m_{\beta}) & = & \frac{N_{\alpha}!N_{\beta}!}{2^{M}N!m_{1}!\cdots m_{4}!m_{\alpha}!m_{\beta}!}\int_{-\pi}^{\pi}\frac{d\lambda}{2\pi}\int_{-\pi}^{\pi}\frac{d\Lambda}{2\pi}\nonumber \\
 &  & \times\cos\left[(N_{\alpha}-N_{\beta}-m_{\alpha}+m_{\beta})\Lambda\right]\nonumber \\
 &  & \times\prod_{i=1}^{4}\left[\cos\Lambda+\eta_{i}\cos\left(\lambda+\varphi_{i}\right)\right]^{m_{i}}\label{eq:2InterfPO}\end{eqnarray}
 where $M=m_{1}+\cdots+m_{4},$ $\eta_{1}=\eta_{3}=1$; $\eta_{2}=\eta_{4}=-1$;
$\varphi_{1}=\varphi_{2}=-\zeta$; $\varphi_{3}=\varphi_{4}=\theta$.
This result is essentially the same as Eq.\ (\ref{OldProb}) with
the addition now of the cosine transform in $\Lambda.$ This factor
allows population oscillations as in Sec. IV. However now we have
the option of adjusting relative phases between the two interferometer
sets.

A summation version of the probability is much more convenient for
computing population oscillations. A result analogous to Eq. (\ref{Prob2Cond})
of Sec. \ref{Populationoscillations} is \begin{eqnarray}
\mathcal{P}(m_{1},m_{2},m_{3},m_{4},m_{\alpha},m_{\beta}) & = & \frac{m_{1}!\cdots m_{4}!}{m_{\alpha}!m_{\beta}!2^{N+2M}}\sum_{p_{2}p_{3}p_{4}}\frac{(-1)^{p_{2}+p_{4}}e^{-i(\zeta+\theta)(p_{3}+p_{4})}}{p_{2}!p_{3}!p_{4}!(m_{2}-p_{2})!(m_{3}-p_{3})!(m_{4}-p_{4})!}\nonumber \\
 &  & \times\frac{1}{(N_{\alpha}-m_{\alpha}-p_{2}-p_{3}-p_{4})!}\nonumber \\
 &  & \times\frac{1}{(m_{1}-N_{\alpha}+m_{\alpha}+p_{2}+p_{3}+p_{4})!}\label{eq:2InterfPOSum}\end{eqnarray}

\begin{figure}[h]
 \includegraphics[width=3in]{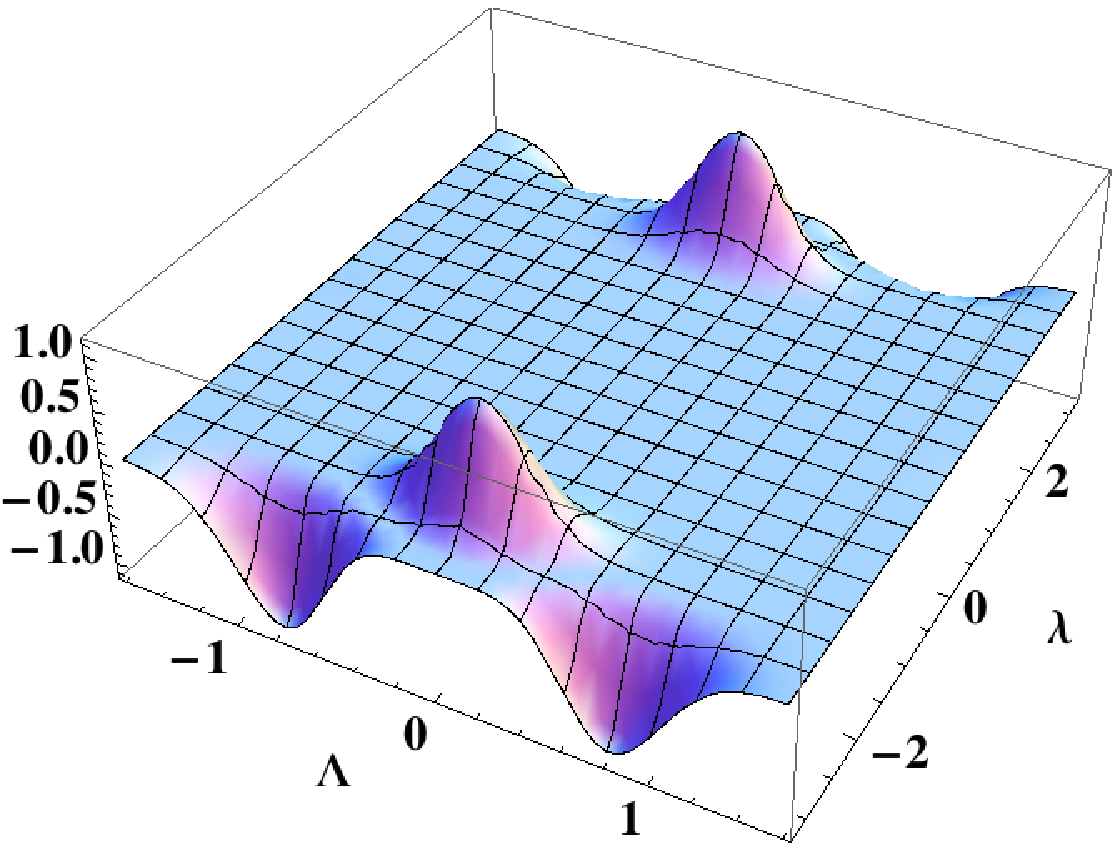}\hspace{0.2in}
\includegraphics[width=3in]{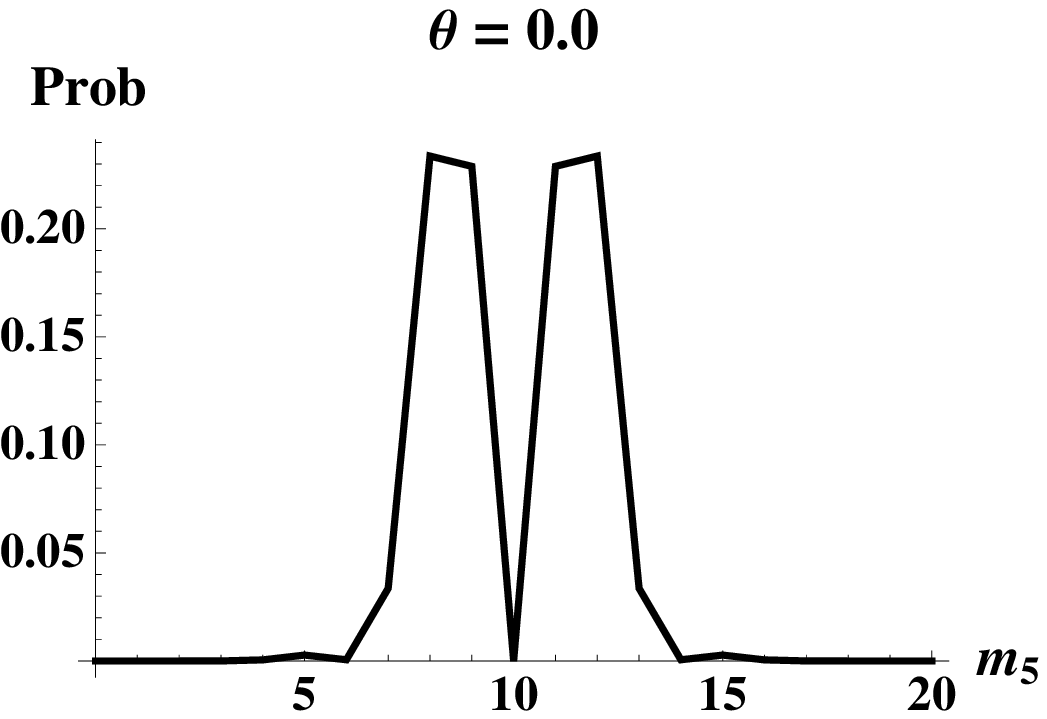}

\caption{(Color online) Double interferometer results for $\{m_{1},m_{2},m_{3},m_{4}\}=\{$2,8,1,9\}
with phase shifts $\zeta=\theta=0$. Left: The last line of the
integrand of Eq.\ (\ref{eq:2InterfPO}). There are two equal classical
peaks (on $\Lambda=0$ axis) and negative quantum peaks. Right. The
corresponding population oscillations with a {}``dark fringe'' at
the center.}

\label{TwoInterfZeroAngles} 
\end{figure}

\begin{figure}[h]
\includegraphics[width=3in]{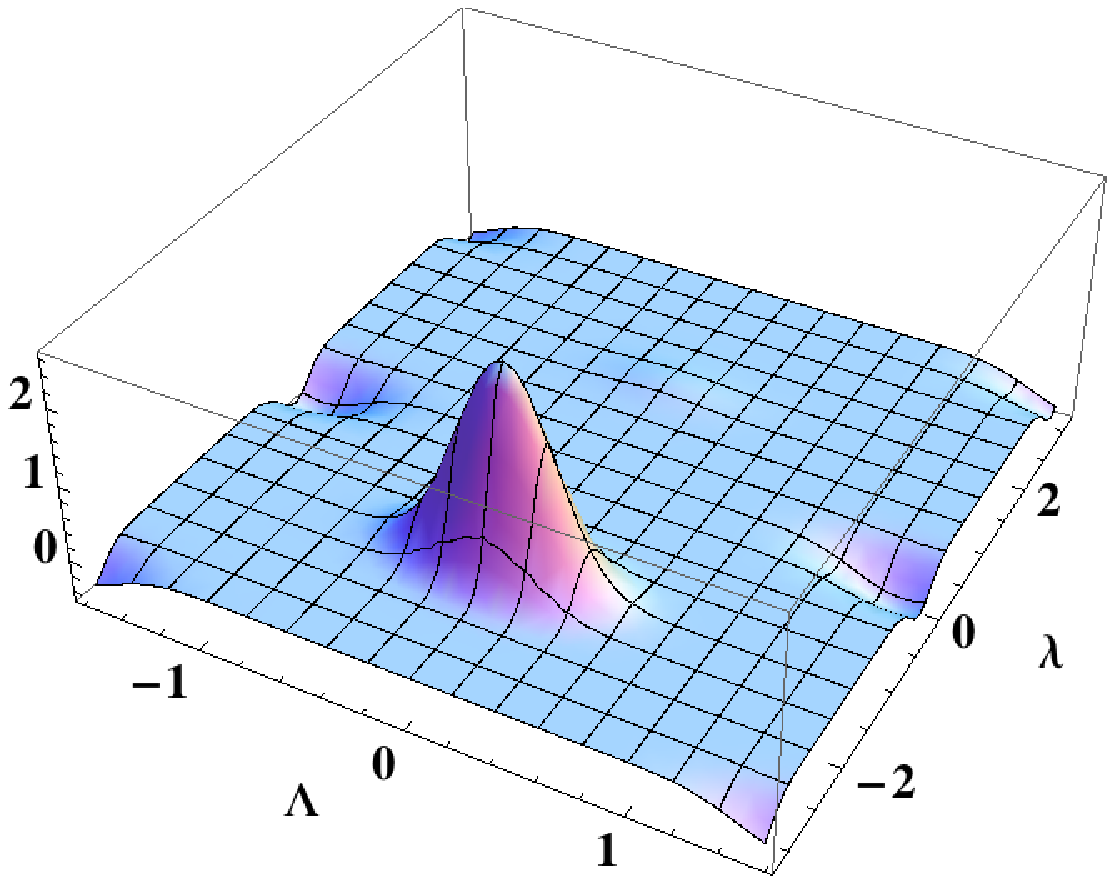} \hspace{0.2in}
\includegraphics[width=3in]{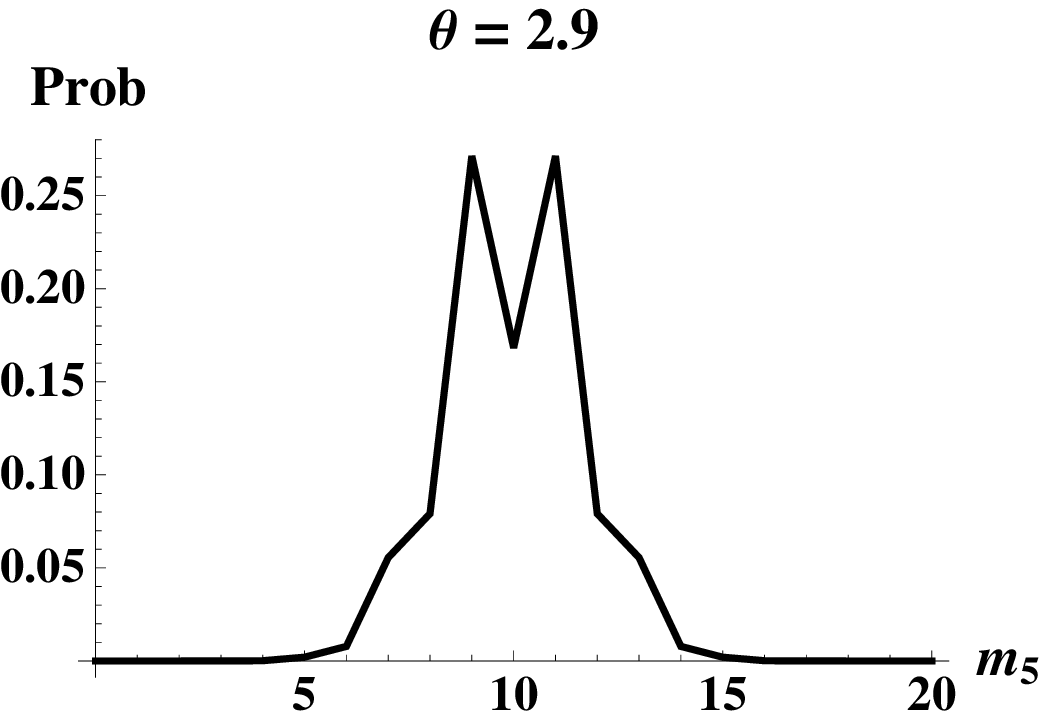}

\caption{(Color online) Double interferometer results for $\{m_{1},m_{2},m_{3},m_{4}\}=\{$2.8,1,9\}
with phase shifts $\zeta=0$, $\theta=2.9.$ Left: One of the classical
peaks has almost completely disappeared and the negative quantum peaks
are now very much smaller. Right: The central depression of the population
oscillation plot no longer goes to zero but becomes an indentation. }

\label{TwoInterfOneNonZeroAngle} 
\end{figure}

Because two independent settings $\theta$ and $\zeta$ are now available,
the phase sign ambiguity can be removed. As a consequence, by adjusting
the phase angles on the interferometers, we can now control the relative
sizes of the two classical peaks, i. e., those along $\Lambda=0$.
Consider the following plots where we show the last line of the integrand
of Eq.\ (\ref{eq:2InterfPO}) and the population oscillations given
by Eq.\ (\ref{eq:2InterfPOSum}) associated with the same parameters.
With the phase shifters set at zero the two classical peaks have equal
sizes and there is a definite population oscillation structure (Fig.\ \ref{TwoInterfZeroAngles}).
However, with a different phase shift, one of the classical peaks
can be made much smaller as seen in Fig.\ \ref{TwoInterfOneNonZeroAngle}
and the quantum peaks become smaller as well. Moreover, the population
oscillation central zero no longer vanishes. For other phase shift
angles (for instance $\zeta=0$ and $\theta=2.5$ radian), the integrand
can be reduced to a single classical peak with no peak in the quantum
region at all; the corresponding population oscillation central dip
then becomes a simple peak.

If the state vector is the sum of two components centered around two
different values of the phase, and if the norm of one component is
larger than that of the other, one obtains two peaks in the classical
region, one large and one small. The small classical peak corresponds
to a population in the phase representation, so that it is second
order with respect to the second component of the state vector. By
contrast, the peaks in the quantum region are first order, since they
correspond to off-diagonal matrix elements. As a consequence, when
one reduces the small phase component, the small classical peak disappears
more rapidly than the quantum peaks. This explains why the left of
Fig.\ \ref{TwoInterfOneNonZeroAngle} has a classical peak that is
barely visible, but still clearly shows the (negative) quantum peaks.

\end{document}